\documentclass[twocolumn,superscriptaddress,showpacs,nofootinbib,preprintnumbers,secnumarabic,amssymb, nobibnotes, aps, prd]{revtex4-2}
\usepackage[utf8]{inputenc}
\usepackage{quantikz}
\usepackage{graphicx} 
\usepackage{subcaption}
\usepackage{latexsym,amsmath,amssymb,amsthm,lmodern,float,url}
\usepackage{natbib}
\usepackage{color}
\usepackage{microtype}
\usepackage{import}
\usepackage{bbold}
\usepackage[plain]{fancyref}
\usepackage{varioref}
\usepackage{slashed}
\usepackage{multirow}
\usepackage{tikz}
\usepackage{scrextend}
\usepackage{braket}
\usepackage{units}
\usetikzlibrary{shapes}
\usetikzlibrary{positioning}
\usepackage[normalem]{ulem}
\usepackage{amsmath}  
\usepackage{bm}
\usepackage{caption}
\usepackage{ragged2e}
\DeclareCaptionJustification{justified}{\justifying}
\newcommand{\fig}[1]{Fig.~\ref{fig:#1}}

\newcommand{\eq}[1]{Eq.~(\ref{eq:#1})}

\definecolor{seagreen}{rgb}{0.180392,0.545098,0.341176}

\usepackage[colorlinks=true,backref=false, linktocpage=true,
citecolor=blue,urlcolor=blue,linkcolor=blue,pdfpagemode=UseOutlines]{hyperref}
\hypersetup{
  bookmarksnumbered=true,
  pdftitle = {},
  pdfsubject = {},
  pdfauthor = {},
  pdfkeywords = {}
}

\begin{document}

\title{Hardware-efficient quantum simulation of intense-field QED}

\author{Zhuoyi Li}
\thanks{Contributed equally to this work.}
\affiliation{International Centre for Theoretical Physics Asia-Pacific, University of Chinese Academy of Sciences, 100190 Beijing, China}
\affiliation{School of Physical Sciences, University of Chinese Academy of Sciences, Beijing 100049, China}
\author{Bin Xu}
\thanks{Contributed equally to this work.\\
\href{mailto:binxu@kias.re.kr}{binxu@kias.re.kr}}

\affiliation{School of Physics and State Key Laboratory of Nuclear Physics and Technology, 
	Peking University, Beijing 100871, China}
\affiliation{School of Physics, Korea Institute for Advanced Study, Seoul, 02455, Republic of Korea}

\author{Zhongtian Dong}
\affiliation{Center for High Energy Physics, Peking University, Beijing 100871, China}
\author{Yuxiang Huang}
\affiliation{Laboratory of Spin Magnetic Resonance, School of Physical Sciences, Anhui Province Key Laboratory of Scientific Instrument Development and Application, University of Science and Technology of China, Hefei, 230026, China}

\author{Ying-Ying Li}
\email{liyingying@ihep.ac.cn}
\affiliation{Institute of High Energy Physics, Chinese Academy of Sciences, Beijing 100049, China}

\author{Yiheng Lin}
\affiliation{Laboratory of Spin Magnetic Resonance, School of Physical Sciences, Anhui Province Key Laboratory of Scientific Instrument Development and Application, University of Science and Technology of China, Hefei, 230026, China}
\affiliation{Hefei National Laboratory, University of Science and Technology of China, Hefei, 230088, China}
\affiliation{Hefei National Research Center for Physical Sciences at the Microscale,
University of Science and Technology of China, Hefei 230026, China}

\author{Jing Shu}
\email{jshu@pku.edu.cn}
\affiliation{School of Physics and State Key Laboratory of Nuclear Physics and Technology, 
	Peking University, Beijing 100871, China}
\affiliation{Center for High Energy Physics, Peking University, Beijing 100871, China}
\affiliation{Beijing Laser Acceleration Innovation Center, Huairou, Beijing, 101400, China}

\date{\today}

\begin{abstract}
{Strong electromagnetic backgrounds make quantum electrodynamics a real-time nonperturbative problem involving dressed fermions and dynamical photons. We propose a trapped-ion protocol for simulating intense-field QED in $3+1$ dimensions in the Furry picture. The construction encodes photon modes in collective phonons and Volkov-dressed fermion modes in ion spins, combining native spin-phonon couplings with Clifford circuits that compress nonlocal Jordan--Wigner strings. For nonlinear Breit--Wheeler pair production, the protocol has polynomial resource scaling and is benchmarked against exact single-mode dynamics with controlled Trotter errors. With experimentally motivated phonon heating and dephasing, zero-noise extrapolation substantially reduces deviations in photon-survival and pair-production signals. These results provide a hardware-efficient route to intense-field particle-production dynamics beyond perturbative or static-field descriptions.}
\end{abstract}

\maketitle

\textbf{\textit{Introduction -- }}
{Quantum electrodynamics in ultra-strong electromagnetic backgrounds is a paradigmatic arena for real-time nonperturbative quantum dynamics. Treating the background field to all orders gives rise to processes such as nonlinear Breit--Wheeler pair production and nonlinear Compton scattering, which are central to high-intensity laser experiments~\cite{Fedotov:2022ely, DiPiazza:2011tq} and to QED in extreme astrophysical environments~\cite{Kaspi:2017fwg}. A fully quantum description of these processes must retain background-field dressing, dynamical photon modes, and fermionic degrees of freedom simultaneously. This combination makes intense-field QED (IFQED) a sharply defined target for quantum simulation and a setting in which a hardware protocol could connect high-energy physics, quantum information, and atomic physics.}

Quantum simulation has become a promising route to nonperturbative real-time dynamics in quantum field theory~\cite{Banuls:2019lgt, Bauer:2022hpo, Funcke:2023lftreview, DiMeglio:2023nsa, Fang:2024ple}. Existing digital approaches to IFQED~\cite{Hidalgo:2023wzr, Draper:2025nku} generally encode photon modes into qubits after truncating their occupation levels, increasing the resource cost of bosonic dynamics. Trapped ions offer a complementary possibility because they provide controllable spin degrees of freedom together with quantized collective motion~\cite{Davoudi:2021ney}. A concrete route is still needed from the Furry-picture formulation of IFQED~\cite{Furry:1951bef} in \(3+1\) dimensions to trapped-ion operations, with dynamical photons represented by hardware-native bosons and nonlocal fermionic correlations implemented efficiently.

\begin{figure*}
\centering
\includegraphics[width=\textwidth]{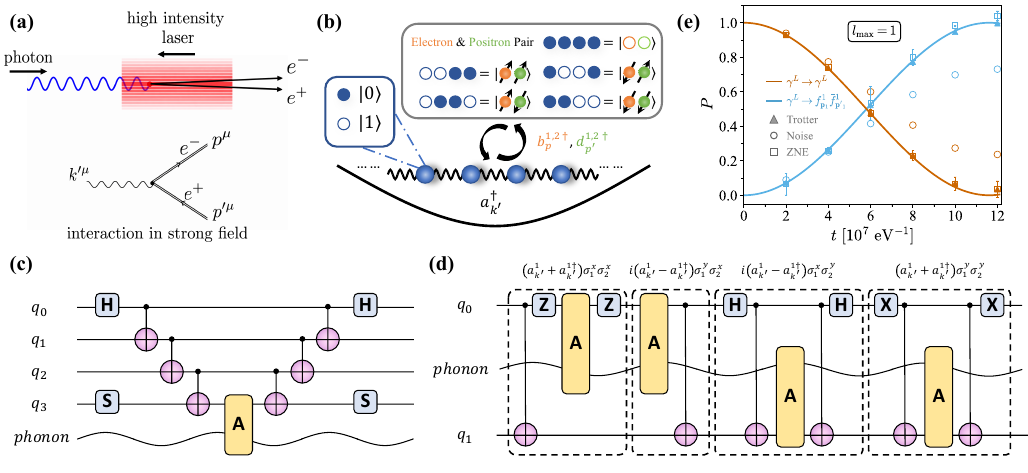}
\caption{\label{fig:integrate-f}
	{$\bm{(a)}$ Nonlinear Breit--Wheeler pair production from a high-energy photon interacting with a strong electromagnetic background, together with the corresponding Feynman diagram.
	$\bm{(b)}$ Mapping of IFQED degrees of freedom onto a trapped-ion system.
	$\bm{(c)}$ Hybrid analog--digital circuit implementing $e^{-i\theta H_{ij}}$ for the interaction $H_{ij}$ in \eq{hij}, shown for $i=0$ and $j=3$. Gate $\mathbf{A}$ denotes the analog spin-phonon gate that implements the local boson-fermion interaction.
	$\bm{(d)}$ Hybrid analog--digital circuit for the single-mode interaction term $a^{L\dagger}_{k'} d^1_{p'} b^1_{p}+\mathrm{h.c.}$
	$\bm{(e)}$ Dynamical benchmark. For $t\leq 4\times10^7\mathrm{eV}^{-1}$, one Trotter step is used; for $t>4\times10^7\mathrm{eV}^{-1}$, two Trotter steps are used to keep Trotter errors small. Orange and Cyan denote the probability of the survival channel $\gamma^L\to\gamma^L$ and the pair-production channel $\gamma^L\to f_{\mathbf{p}_1}^1\bar f_{\mathbf{p}'_1}^1$, respectively. Solid lines show exact time evolution under Eq.~\eqref{eq:subHam}; triangles denote noiseless trotterized circuit simulations; open circles denote noisy simulations of the same circuit; and squares show zero-noise extrapolation (ZNE) results. Error bars indicate the fitting uncertainty of the noise extrapolation and are enlarged by a factor of fifteen for visibility.}}
\end{figure*}

Here we construct such a route. Photon modes are encoded into collective phonons, while Volkov-dressed fermionic modes are mapped to ion spins by a Jordan--Wigner transformation~\cite{Jordan:1928wi}. The central technical step is a hybrid analog--digital compilation: Clifford circuits compress Jordan--Wigner strings, and native spin-phonon gates implement the resulting local boson-fermion couplings. The number of digital CNOT gates per Trotter step scales as \(\mathcal{O}(N_p^3)\), and the number of analog spin-phonon gates scales as \(\mathcal{O}(N_p^2)\), where \(N_p\) is the number of retained momentum modes. As a benchmark, we simulate nonlinear Breit--Wheeler pair production, including harmonic truncation, Trotterization, trapped-ion noise, and zero-noise extrapolation. Derivations of the IFQED Hamiltonian, the trapped-ion spin-phonon implementation, the multimode extension, and the noise model are provided in the Supplemental Material. The overall setup and implementation scheme are illustrated in \fig{integrate-f}.

\textbf{\textit{IFQED framework and trapped-ion mapping -- }}
We now detail the formulation of the IFQED Hamiltonian and its encoding on a trapped-ion platform. Throughout, the laser field $\mathcal{A}^\mu$ is treated as a classical background and modeled as a left-circularly polarized plane wave. Fermions propagating in this background are described in the Furry picture by Volkov states $\psi^{(\pm)s}_{p}(x)$, which resum the interaction with the background field to all orders. Here, the superscript $\pm$ distinguishes particles from antiparticles, while $s=1,2$ labels the spin degree of freedom. By contrast, the quantized photon field is kept in its free form, and the effect of the background field enters through the exchange of an effective harmonic number $l$. {Details of the Furry-picture construction and the derivation of the interaction Hamiltonian are given in the IFQED Furry-picture section of the Supplemental Material.}

To streamline the notation, we group the photon and fermionic operators into quadruplets. For a photon of momentum $\mathbf{k}'$ with polarization $r=L,R$, we define
\begin{align}
	\hat{A}^{n}_{\mathbf{k}'} &=
	\left\{
	a^{L}_{\mathbf{k}'},
	a^{L\dagger}_{\mathbf{k}'},
	a^{R}_{\mathbf{k}'},
	a^{R\dagger}_{\mathbf{k}'}
	\right\},
\end{align}
and for fermionic Volkov modes of momentum $\mathbf{p}$,
\begin{align}
	\hat{B}^{m}_{\mathbf{p}} &=
	\left\{
	b^{1}_{\mathbf{p}},
	d^{1\dagger}_{\mathbf{p}},
	b^{2}_{\mathbf{p}}, d^{2\dagger}_{\mathbf{p}}
	\right\}, 
\end{align}
with $n,m=1,2,3,4$, where $b^s_{\mathbf{p}}$ annihilating an electron and $d^{s\dagger}_{\mathbf{p}}$ creating a positron. After discretizing momentum space on a lattice of linear size $L$ with momentum spacing $\Delta p = 2\pi/L$, the interaction Hamiltonian takes the compact form
\begin{align}
	H_{\mathrm{int}}
	=
	\sum_{\mathbf{k}'\,\mathbf{p}'\,\mathbf{p}}
	\sum_{n m' m}
	\sum_{l=-\infty}^{\infty}
	c_{l}^{n m' m}\,
	\hat{A}^{n}_{\mathbf{k}'}
	\hat{B}^{m'\dagger}_{\mathbf{p}'}
	\hat{B}^{m}_{\mathbf{p}}\,
	e^{iQ^{0}t}\,
	\delta^{(3)}(\mathbf{Q}).
	\label{eq:H_BW}
\end{align}
Here, $Q \equiv \{Q^0, \mathbf{Q}\} = l k + (-1)^n k' - (-1)^{m'} q' + (-1)^m q$, 
where the fermionic quasi-momenta are $q_{\mu} = p_{\mu} + \frac{e^{2}a^{2}}{2\,p\!\cdot\! k}\,k_{\mu}$ and $q'_{\mu} = p'_{\mu} + \frac{e^{2}a^{2}}{2\,p'\!\cdot\! k}\,k_{\mu}$.
The spatial delta function $\delta^{(3)}(\mathbf{Q})$ enforces momentum conservation, while the phase factor $e^{iQ^{0}t}$ governs the real-time oscillatory dynamics. The coefficients $c_{l}^{n m' m}$ encode the effective couplings between one photon mode and a fermion-antifermion pair, including their dependence on momentum, polarization, spin, and the harmonic index $l$.

For a $(d+1)$-dimensional momentum lattice with $N$ retained modes in each spatial direction, the total number of momentum modes is $N_p=N^d$. In what follows we focus on the physically relevant case $d=3$.

The mapping to trapped-ion hardware is illustrated schematically in \fig{integrate-f}($\bm{b}$). Photon mode $a^r_{\mathbf{k}'}$ is encoded directly into a collective phonon mode labeled $a^r_{k}$, while the fermionic degrees of freedom are represented by ion spins through a Jordan--Wigner transformation. We order the fermionic modes sequentially along a one-dimensional spin chain, with the four internal states associated with each momentum mode occupying four consecutive qubits. This yields
\begin{align}
	b_{\mathbf{p}_n}^{1\dagger}
	&=
	\prod_{i<4n}\sigma_i^z\,\sigma_{4n}^{+},
	&
	d_{\mathbf{p}_n}^{1\dagger}
	&=
	\prod_{i<4n+1}\sigma_i^z\,\sigma_{4n+1}^{+},
	\nonumber\\
	b_{\mathbf{p}_n}^{2\dagger}
	&=
	\prod_{i<4n+2}\sigma_i^z\,\sigma_{4n+2}^{+},
	&
	d_{\mathbf{p}_n}^{2\dagger}
	&=
	\prod_{i<4n+3}\sigma_i^z\,\sigma_{4n+3}^{+}.
\end{align}
Under this encoding, representative interaction terms satisfying energy-momentum conservation for the photon-induced pair creation take the form
\begin{equation}
	H_{\mathrm{int}}
	\supset
	c^r_{kij}\,a^r_{k}\,\sigma_i^{+}\Sigma^{z}_{ij}\sigma_j^{+}
	+ \mathrm{h.c.},
	\label{eq:Hion}
\end{equation}
where $c^r_{kij}$ is the coefficient for the interaction between the phonon mode $a^r_k$ and spins $i$ and $j$. Energy conservation requires $Q^0 = 0$, so the phase factor in Eq.~\eqref{eq:H_BW} becomes unity for the resonant terms considered here.
Equation~\eqref{eq:H_BW} also contains nonlinear Compton scattering terms, which take the form $a^r_{k}\,\sigma_i^{-}\Sigma^{z}_{ij}\sigma_j^{+} + \mathrm{h.c.}$ The Jordan--Wigner string $\Sigma^{z}_{ij}\equiv \prod_{i<m<j}\sigma_m^{z}$ reflects the nonlocal fermionic structure generated by the encoding; handling these strings efficiently is the central reason for adopting the hybrid analog--digital strategy developed in the next section.

\textbf{\textit{Hybrid analog-digital implementation and scaling-- }}
To realize the nonlocal spin strings coupled to bosonic phonon modes, we adopt a hybrid analog-digital strategy tailored to trapped-ion hardware. The key idea is to use digital gate sequences to compress the nonlocal Jordan--Wigner strings into local operators, while exploiting native spin-phonon interactions to implement the resulting boson-fermion couplings efficiently.

We now describe the time evolution generated by the interaction terms in $H_{\mathrm{int}}$. To systematically reduce the length of the Jordan--Wigner strings appearing in Eq.~\eqref{eq:Hion}, we employ digital gate constructions based on purely Clifford gates, including nearest-neighbor CNOT gates, the Hadamard gate $H$ and the phase gate $S$.
The basic relation is
\begin{equation}
	\sigma_i^{\alpha}\sigma_{i+1}^{\beta}
	=
	U_i^{\alpha\dagger}\,\sigma_{i+1}^{\beta}\,U_i^{\alpha},
	\qquad
	\alpha\in\{x,y,z\},\ \beta\in\{y,z\},
\end{equation}
where $U_i^\alpha$ acts on qubits $i$ and $i+1$. Explicitly,
\begin{align}
	U_i^x &= \mathrm{CNOT}_{i,i+1}\,H_i, \notag\\
	U_i^y &= \mathrm{CNOT}_{i,i+1}\,H_i S_i H_i, \\
	U_i^z &= \mathrm{CNOT}_{i,i+1},\notag
	\label{eq:Ualpha}
\end{align}

As a representative example, we consider an interaction term of the form
\begin{equation}
\label{eq:hij}
	H_{ij}
	=
	c^r_{kij}\, a^r_{k}\,\sigma_i^{x}\Sigma^z_{ij}\sigma_j^{x}
	+ \mathrm{h.c.}.
\end{equation}
To map the system into the appropriate trapped-ion spin–phonon interaction, we first apply a rotation operator on site $j$, i.e. $S_j^{\dagger} \sigma^x_j S_j=\sigma_j^y$. The time evolution generated by $H_{ij}$ can then be written as
\begin{align}
	e^{-i H_{ij} t}
	&=
	S_j^{\dagger}
	e^{-i\theta\left(a^r_{k}\,\sigma_i^x \Sigma^z_{ij}\sigma_j^x+\mathrm{h.c.}\right)}S_j
	\nonumber\\
	&=
	S_j^{\dagger} U_i^{x\dagger} U_{i+1}^{z\dagger}\cdots U_{j-1}^{z\dagger}
	e^{-i\theta\left(a^r_{k}+a_k^{r\dagger}\right)\sigma_j^y}\nonumber\\
	&\quad\times U_{j-1}^{z}\cdots U_{i+1}^{z} U_i^{x}S_j.
    \label{eq:uhij}
\end{align}
with $\theta$ being the product of the evolution time $t$ and the coupling strength in the Hamiltonian. Thus, through a recursive application of these local unitaries, a nonlocal spin string is reduced to an effective local coupling between the phonon mode and a single spin at site $j$. 

Given native ion-laser couplings of the form~\cite{PhysRevResearch.3.043072,monroe2021programmable}
\begin{equation}
	H_{k,j}^{\sigma a}(\phi_{k,j})
	=
	\frac{\eta_{k,j}\Omega_j}{2}
	\left(
	a^r_k e^{i\phi_{k,j}}
	+
	a_k^{r\dagger} e^{-i\phi_{k,j}}
	\right)\sigma_j^y,
	\label{eq:laser-ion}
\end{equation}
where $\eta_{k,j}$ is the Lamb--Dicke parameter and $\Omega_j$ is the Rabi frequency, the analog spin-phonon gate can realize the following unitary operation
\begin{align}
	R_{k,j}^{\sigma a}(\theta_{k,j},0) \equiv e^{- i H_{k,j}^{\sigma a}(0)\tau_{gate}}
	&=e^{-i\theta_{k,j}(a^r_k+a_k^{r\dagger})\sigma_j^y},
	\notag\\
	R_{k,j}^{\sigma a}\!\left(\theta_{k,j},\frac{\pi}{2}\right)\equiv e^{-i H_{k,j}^{\sigma a}(\frac{\pi}{2})\tau_{gate}}
	&=e^{\theta_{k,j}(a^r_k-a_k^{r\dagger})\sigma_j^y}.
	\label{eq:analog_gate}
\end{align}
Further implementation details are given in the spin-phonon Hamiltonian section of the Supplemental Material.

With $\theta_{k,j}= \frac{\eta_{k,j}\Omega_j}{2}\tau_{gate}$ matched to the $\theta$ parameter in \eq{uhij}, the local coupling between the phonon mode and the spin at site $j$ can be implemented. In this way, the digital part of the protocol removes the nonlocal fermionic strings, while the analog part implements the native spin-phonon dynamics. A concrete circuit example for $j=i+3$ is shown in \fig{integrate-f}($\bm{c}$), where the analog spin-phonon gate is denoted by $\mathbf{A}$.

Using these ingredients, the time evolution under the full Hamiltonian can be constructed via a Trotter--Suzuki decomposition \cite{Suzuki:1976be}. For an interaction term with spin-string length $m$, the digital reduction requires $2(m-1)$ CNOT gates and has the same circuit depth. Since the maximal string length is of order $N_p$, and the number of noncommuting interaction terms scales as $\mathcal{O}(N_p^2)$, one Trotter step requires $\mathcal{O}(N_p^3)$ CNOT gates. In addition, implementing the boson-fermion couplings requires $\mathcal{O}(N_p^2)$ analog spin-phonon gates per Trotter step. For a total evolution time $t$, the number of first-order Trotter steps needed to achieve error $\epsilon$ scales as $N_t \sim \mathcal{O}\!\left(t^2 N_p^2 C/\epsilon\right)$,
where $C$ denotes a typical magnitude of $\|[H_{ij},H_{i'j'}]\|$ among noncommuting terms. The resulting CNOT depth therefore scales as $\mathcal{O}\!\left(N_p^5 t^2 C/\epsilon\right)$ and the depth of the analog spin-phonon gates scales as $\mathcal{O}\!\left(N_p^4 t^2 C/\epsilon\right)$.
Since the digital component consists entirely of Clifford gates, the nonclassical computational power of the protocol is expected to arise from the analog spin-phonon operations.

\textbf{\textit{Benchmark of single-mode nonlinear Breit--Wheeler dynamics -- }}
As a proof-of-principle benchmark of our protocol, we consider a minimal single-mode realization of nonlinear Breit--Wheeler pair production. {Specifically, we restrict the system to one photon mode with momentum $k'^\mu$, one electron mode with momentum $p_1^\mu$, and one positron mode with momentum $p_1'{}^\mu$.}

We choose a left-circularly polarized background laser field propagating along the negative $z$ direction, with four-momentum $k^\mu=(\omega,0,0,-\omega)$,
and an incident photon propagating along the positive $z$ direction, with four-momentum $k'^\mu=(\omega',0,0,\omega')$. For the benchmark calculation, we adopt $\omega = 1.55~\mathrm{eV}$, $\omega' = 1~\mathrm{TeV}$, and the laser intensity parameter $\xi=\frac{e a}{m_e}=1$, where $a=|\mathcal{A}_\mu|$ is the amplitude of the background potential. For these parameters, the quantum nonlinearity parameter $\chi=\frac{k\cdot k'}{m_e^2}\xi =11.87$, placing the process in a regime where IFQED effects are significant.

For the purpose of demonstration, we first consider including only the $l = 1$ term. 
Assuming equal momenta along the $z$ direction for the produced fermion and antifermion, quasi-momentum conservation $q_1+q_1'=k+k'$ yields
\begin{align}
	q_1^0 &= q'_1{}^0 = \frac{\omega'+\omega}{2} \simeq 0.5~\mathrm{TeV}, \notag\\
	\mathbf{q}_1^3 &= \mathbf{q}'_1{}^{3} = \frac{\omega'-\omega}{2} \simeq 0.5~\mathrm{TeV}, \notag\\
	|\mathbf{q}_1^\perp|
	&= |\mathbf{q}'_1{}^{\perp}|
	= \sqrt{\omega\omega'-m_*^2}
	= 1.014~\mathrm{MeV}.
	\label{eq:num2}
\end{align}
where $m_*^2=q_1^2=m_e^2(1+ \xi^2)$ is the electron effective mass in the field.
{Since the characteristic energy scale of the process is of order TeV, while different momentum modes should remain distinguishable at the MeV scale, we take the discretized momentum spacing to be $\Delta p = 0.03~\mathrm{MeV}$. This corresponds to $L=\frac{2\pi}{\Delta p}=4.12\times10^{-11}\,\mathrm{m}$.}

Under these assumptions, the interaction Hamiltonian reduces to six dominant terms,
\begin{align}
	H_{\mathrm{int}}=&(-1.35 a_{\mathbf{k'}}^{L \dagger}d^1_{\mathbf{p'}} b^1_{\mathbf{p}}
	+3.23 a_{\mathbf{k'}}^{L \dagger}d^1_{\mathbf{p'}} b^2_{\mathbf{p}}
	+1.35 a_{\mathbf{k'}}^{L \dagger}d^2_{\mathbf{p'}} b^2_{\mathbf{p}}\nonumber\notag\\
	&-2.93 a_{\mathbf{k'}}^{R \dagger} d^1_{\mathbf{p'}} b^1_{\mathbf{p}}
	-3.23 a_{\mathbf{k'}}^{R \dagger}d^2_{\mathbf{p'}} b^1_{\mathbf{p}}
	+2.93 a_{\mathbf{k'}}^{R \dagger} d^2_{\mathbf{p'}} b^2_{\mathbf{p}})\nonumber\notag\\
	&\times 10^{-8} \mathrm{eV} + \mathrm{h.c.},  
	\label{eq:Hint}
\end{align}
where the asymmetry between the coefficients of the left- and right-polarized photon sectors arises from the left-circular polarization of the background field.

\begin{figure*}[htbp]
    \centering
    \begin{subfigure}[c]{0.328\textwidth}
        \centering
        \includegraphics[width=\textwidth]{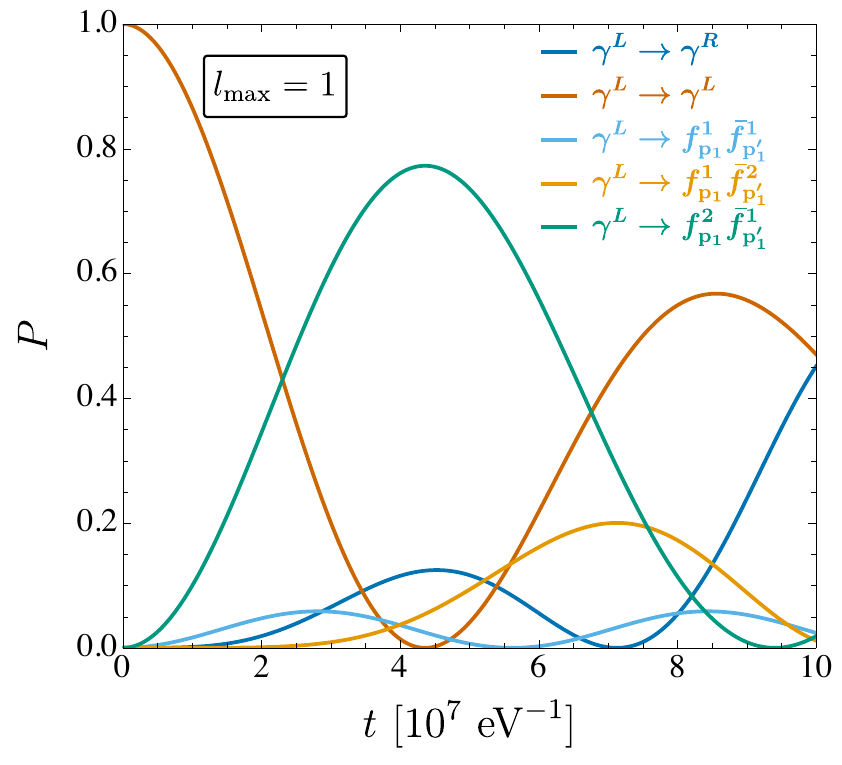}
        \caption{}  
        \label{fig:2a}
    \end{subfigure}
    \hfill
    \begin{subfigure}[c]{0.328\textwidth}
        \centering
        \includegraphics[width=\textwidth]{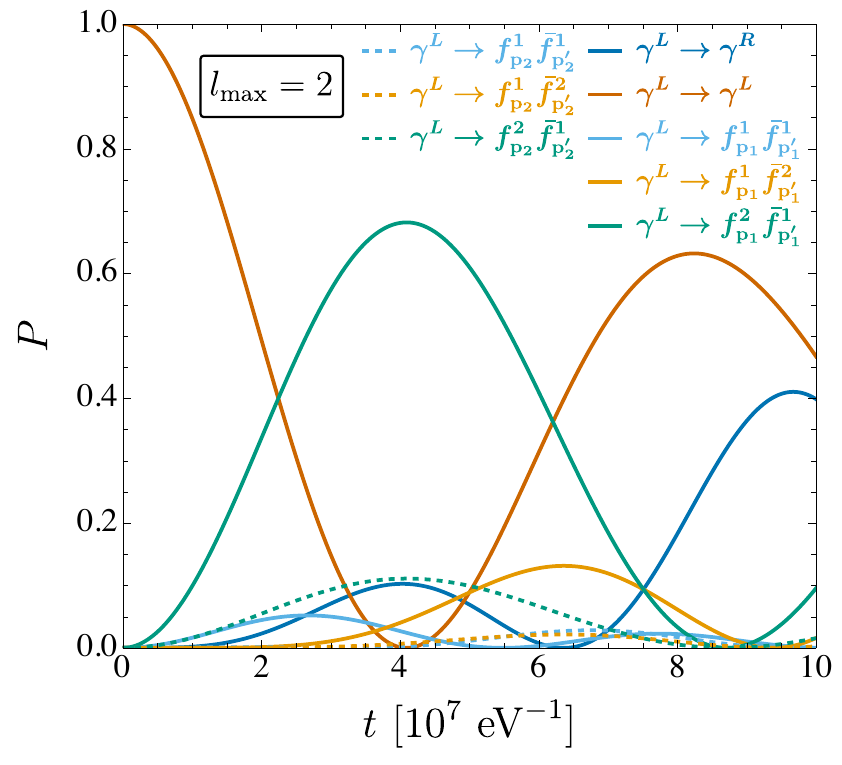}
        \caption{} 
        \label{fig:2b}
    \end{subfigure}
    \hfill
    \begin{subfigure}[c]{0.328\textwidth}
        \centering
        \includegraphics[width=\textwidth]{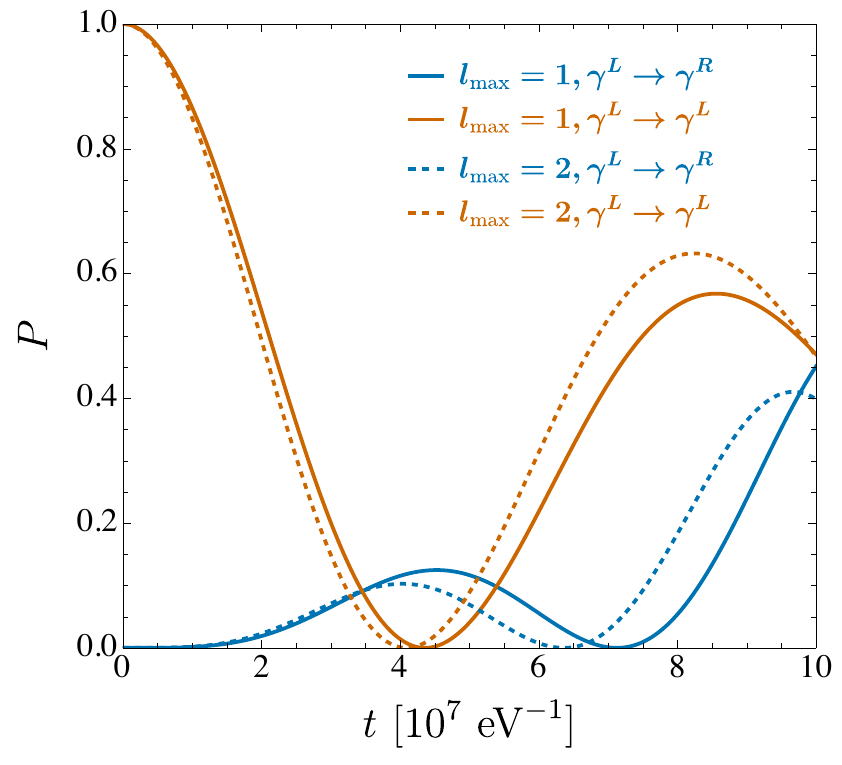}
        \caption{} 
        \label{fig:2c}
    \end{subfigure}

    \caption{
    Time evolution of channel probabilities for an incident left-polarized photon.
    The probabilities for $\gamma^L \to f_{\mathbf{p}_1}^1 \bar{f}_{\mathbf{p}'_1}^{1}$ and $\gamma^L \to f_{\mathbf{p}_1}^2 \bar{f}_{\mathbf{p}'_1}^{2}$ are identical, as are those for $\gamma^L \to f_{\mathbf{p}_2}^1 \bar{f}_{\mathbf{p}'_2}^{1}$ and $\gamma^L \to f_{\mathbf{p}_2}^2 \bar{f}_{\mathbf{p}'_2}^{2}$; each pair is therefore shown with the same color.
    (a) Dynamics with harmonic truncation \(l_{\max}=1\).
    (b) Dynamics with harmonic truncation \(l_{\max}=2\), where solid lines denote channels in the \(l=1\) sector and dashed lines denote channels in the \(l=2\) sector.
    (c) Photon dynamics for the two truncation schemes: solid lines correspond to \(l_{\max}=1\), while dashed lines correspond to \(l_{\max}=2\).}
    \label{fig:evo}
\end{figure*}

We take the initial state to be a left-polarized photon state $|\gamma^L\rangle=a_{\mathbf{k}'}^{L\dagger}|\Omega\rangle$,
where the vacuum state $|\Omega\rangle$ satisfies $\hat{a}^r_{\mathbf{k}'}|\Omega\rangle= \hat{b}^{s}_{\mathbf{p}'}|\Omega\rangle= \hat{d}^s_{\mathbf{p}}|\Omega\rangle=0$.
We then compute the probabilities for the channels $\gamma^L \to \gamma^L$, $\gamma^L \to \gamma^R$, and $\gamma^L \to f^s\bar{f}^{s'}$
where $|f^s\bar{f}^{s'}\rangle = b^{s\dagger}_{\mathbf{p}}d^{s'\dagger}_{\mathbf{p'}}|\Omega\rangle$, and the resulting dynamics are shown in \fig{evo}($\bm{a}$).

The dynamics exhibit the decay of the initial photon into electron-positron pairs, with channel-dependent probabilities determined by the corresponding coupling strengths in Eq.~\eqref{eq:Hint}. In addition, a finite transition probability to the right-polarized photon sector appears through the coupled strong-field dynamics. This single-mode benchmark therefore captures the essential polarization dependence and pair-production structure of the nonlinear Breit--Wheeler process, while remaining simple enough for controlled comparison with trotterized and noisy trapped-ion simulations.

{To illustrate the leading correction, \fig{evo}($\bm{b}$) shows the result of extending the truncation from \(l_{\max}=1\) to \(l_{\max}=2\), where an additional electron mode and positron mode with momenta $p_2^\mu$ and $p_2'{}^\mu$ are included to satisfy quasi-momentum conservation, $q_2+q_2'=l k+k'$, for $l = 2$. The Hamiltonian for the sector with $l = 2$ is given in the multimode-configuration section of the Supplemental Material. The branching ratios into the \(l=2\) channels are suppressed relative to those of the \(l=1\) channels, reflecting the smaller magnitudes of the corresponding coupling coefficients. The dominant decay channel remains \(\gamma^L \to f^2 \bar f^1\) in both truncation schemes. In \fig{evo}($\bm{c}$), we compare the photon dynamics for \(l_{\max}=1\) and \(l_{\max}=2\). The additional $l=2$ channels open further decay pathways, accelerating the decay of the left-polarized photon and causing population in the right-polarized photon sector to appear earlier.}

\textbf{\textit{Noisy simulation and error mitigation --}}
To assess the near-term feasibility of our protocol, we now consider a noisy trapped-ion simulation of the single-mode nonlinear Breit--Wheeler benchmark. For concreteness, we focus on the first term in Eq.~\eqref{eq:Hint}, corresponding to the production of a fermion-antifermion pair with spin configuration $s=s'=1$ from a left-polarized photon mode $a^L_{\mathbf{k}'}$. This photon mode is encoded into the phonon mode $a^L$.
The resulting effective Hamiltonian is
\begin{align}
	H_{\mathrm{int}}^{(1)} = & -1.35\times10^{-8}\,\mathrm{eV} \Big[
	\big(a^{L}+a^{L\dagger}\big)
	\big(\sigma_1^{x}\sigma_2^{x}-\sigma_1^{y}\sigma_2^{y}\big)
	\nonumber\\
	&\qquad
	+\, i\big(a^{L}-a^{L\dagger}\big)
	\big(\sigma_1^{x}\sigma_2^{y}+\sigma_1^{y}\sigma_2^{x}\big)
	\Big],
	\label{eq:subHam}
\end{align}
We take the initial state to be the left-polarized photon and simulate its conversion into a fermion pair under the time evolution generated by Eq.~\eqref{eq:subHam}. The corresponding hybrid analog-digital circuit, obtained by trotterizing the four noncommuting terms in Eq.~\eqref{eq:subHam}, is shown in \fig{integrate-f}($\bm{d}$).

To model a realistic trapped-ion implementation, we incorporate the dominant experimental noise channels affecting both digital and analog operations. We assume characteristic gate durations of $100\,\mu\text{s}$ for CNOT gates and $10\,\mu\text{s}$ for single-qubit gates.
The duration $\tau_{gate}$ of the analog spin-phonon gate $\mathbf{A}$ is adjusted so that $\theta_{k,j} = \theta$ for the chosen evolution time $t$, giving $\tau_{gate} = 10\times (t/10^7 \text{eV}^{-1})\,\mu\text{s}$. These timescales are consistent with high-fidelity operations reported in several trapped-ion platforms, including $^{40}\text{Ca}^+$ systems~\cite{PhysRevLett.127.130505,zhao2025high}, $^{43}\text{Ca}^+$ ions~\cite{PhysRevLett.117.060504}, and $^{9}\text{Be}^+$ ions~\cite{PhysRevLett.117.060505}.
The primary noise sources in our circuit are phonon heating, as well as spin and phonon dephasing. We use a phonon heating rate of $3\,\text{s}^{-1}$, which is within the reach of ion-trap experiments when the stretch mode is used~\cite{PhysRevLett.117.060504,PhysRevLett.117.060505}. We also take phonon and spin coherence times of $3\,\text{ms}$ and $10\,\text{ms}$, respectively. These parameters represent a conservative experimental regime relative to several reported results~\cite{PhysRevLett.117.060504,PhysRevLett.127.130505,PhysRevLett.117.060505,bruzewicz2019trapped}. Gate imperfections from these noise sources are incorporated into the digital and analog operations, with implementation details given in the error-mitigation section of the Supplemental Material. We neglect qubit energy relaxation ($T_1$ processes), whose timescales are typically several orders of magnitude longer than the gate durations considered here~\cite{bruzewicz2019trapped}. We also neglect state-preparation and measurement errors, whose contribution is expected to be subleading in this proof-of-principle analysis.

The initial left-polarized photon state is prepared by starting from the vacuum state $|\Omega\rangle$, realized by initializing all spins in $|0\rangle$, and then exciting the phonon mode corresponding to $a^L$ to a single-phonon state. 
We simulate the time evolution generated by $H^{(1)}_{\rm int}$ using a noisy hybrid analog--digital circuit. 

To mitigate gate noise, we use zero-noise extrapolation (ZNE)~\cite{PhysRevX.8.031027,PhysRevLett.119.180509}. We amplify errors from the two-qubit and analog spin-phonon gates, while leaving single-qubit gates unmitigated because their effect is small in this simulation. Following the standard unitary-folding protocol, a gate $U$ is replaced by $U(U^\dagger U)^n$, which preserves the ideal noiseless evolution while increasing the effective noise exposure. Repeating the simulation for different integers $n$ gives noise-scaled results with folding factor $\mathcal{N}_s = 2n+1$. These data are then extrapolated to the zero-noise limit at $\mathcal{N}_s =0$, as shown in the error-mitigation and zero-noise-extrapolation sections of the Supplemental Material.

The dynamical results are shown in \fig{integrate-f}($\bm{e}$), where we plot the survival channel $\gamma^L\to\gamma^L$ and the pair-production channel $\gamma^L\to f_{\mathbf{p}_1}^1\bar f_{\mathbf{p}'_1}^1$ as functions of time. For $t\leq4\times10^7\mathrm{eV}^{-1}$, we use one Trotter step; for longer times, we use two Trotter steps to keep Trotter errors under control. The agreement between exact evolution and the noiseless trotterized simulation shows that Trotter errors remain small over the plotted time window. With the above noise channels included, visible deviations from the ideal dynamics appear, especially for $t>4\times10^7\mathrm{eV}^{-1}$. ZNE substantially reduces these deviations and approximately recovers the target dynamics. This benchmark indicates that error-mitigated hybrid analog--digital simulation is a viable route toward probing nonperturbative IFQED dynamics on near-term trapped-ion hardware.

\textbf{\textit{Conclusions -- }}
{We have developed a trapped-ion framework for quantum simulation of IFQED in \(3+1\) dimensions. Working in the Furry picture, the protocol encodes dynamical photons in collective phonons and Volkov-dressed fermions in ion spins. A hybrid analog--digital compilation turns nonlocal Jordan--Wigner strings into local spin-phonon couplings using only Clifford gates in the digital layer.}

For nonlinear Breit--Wheeler pair production, the construction is benchmarked against exact single-mode dynamics with negligible Trotter errors and captures the leading effect of an additional harmonic channel.
With experimentally motivated phonon heating and dephasing, noisy circuit simulations show visible deviations from the ideal dynamics, while zero-noise extrapolation approximately recovers both the photon-survival and pair-production observables.

The result is a concrete route for using near-term trapped-ion hardware to access real-time intense-field QED dynamics and a starting point for multimode and multiphoton simulations of nonperturbative particle productions.

\begin{acknowledgments}
We would like to thank Zhongrui Zheng for inspiring discussions. This work was supported by National Natural Science Foundation of China (grant Nos.12450006, 12522509, 12305107, 92565306) and National Key Research and Development Program of China (grant Nos. 2025YFA1614200, 2025YFE0217900). BX is supported in part by a KIAS Individual Grant No. PG103301 at Korea Institute for Advanced Study.
\end{acknowledgments}

\bibliography{draft}

\newpage

\onecolumngrid
\begin{center}
{\Large\bfseries{Supplemental Material}}
\end{center}

\appendix
\section{Intense-field QED in the Furry Picture}
\label{sec:ham}

\subsection{Derivation of Volkov States}

In intense field QED (IFQED), the laser field is usually described as a background field. This back-ground field has to be treated to all orders, which can be achieved by going to the Furry interaction picture in which the background field is treated as part of the unperturbed system.

When an electron interacts with a high-intensity laser, the laser pulse can be described as a plane wave, which we assume to propagate along the negative $z$ axis, and with frequency $\omega = \mathcal{O}(1~\mathrm{eV})$. Maxwells equations in vacuum require that $F_{\mu\nu}$ is transverse, $k_\mu F^{\mu\nu}=0$, where $k_\mu$ is the wave four vector. In that case, the field can be represented by a transverse vector potential, $\mathcal{A}^\mu = (0, \mathbf{A}^{\bot} (\phi), 0)$ satisfying $k \cdot \mathcal{A} = 0$, and parametrized as $\mathbf{A}^{\bot} = \mathcal{A}_0 g(\phi/\Delta\phi)Re[\Vec{\epsilon} e^{-i\phi}]$, where $\phi=k \cdot x$, $\Vec{\epsilon}$ is a complex polarization vector, and $g$ is an envelope function with pulse duration $\Delta \phi$. 

If we take left circular polarization $\Vec{\epsilon}=\frac{1}{\sqrt{2}} \begin{pmatrix}
    1\\-i
\end{pmatrix}$, 
the field of a plane circularly polarized wave can be described by a four-potential $\mathcal{A}^\mu=a_1 \cos{\phi}+a_2 \sin{\phi} $, where $a_{1} , a_{2}$ are the amplitudes of the potential. The vectors $k,a_1$ and $a_2$ are orthogonal to each other, $k\cdot a_1=k\cdot a_2 =a_1 \cdot a_2=0$. Their lengths satisfy $k^2=0, a_1^2=a_2^2=-a^2$. Here we take $a_1=(0,a,0,0)$ and $a_2=(0,0,a,0)$, then $\mathcal{A}^\mu=(0,a \cos{\phi},a \sin{\phi},0)$.

The wave vector should satisfy the light-like four-vector condition $k^2=\omega^2-\mathbf{k}^2=0$, so we assume the wave vector $k^\mu$ of the plane wave is $k^\mu=(\omega,0,0,-\omega)$. The field strength is a univariate function, $F_{\mu\nu} = F_{\mu\nu}(k\cdot x)$ and $\phi = k\cdot x = \omega x^+$ tells us that the field depends solely on the light-front variable $x^+ = t + z$. \cite{Seipt:330321}

Our starting point is the QED Lagrangian for electrons and positrons (Dirac fermion $\psi$) interacting with photons (gauge field $A^{\mu}$), all in the presence of a background electromagnetic field with gauge potential $\mathcal{A}^{\mu}$, $F_{\mu\nu}$ is the field strength of $A^{\mu}$.
\begin{equation}
    \mathcal{L}_{\text{QED}} =  - \frac{1}{4} F^{\mu\nu} F_{\mu\nu}+\bar{\psi} (i \gamma^{\mu} \partial_{\mu} - m) \psi - e\bar{\psi}\gamma^{\mu}\psi (\mathcal{A}_{\mu}+A_{\mu})\label{eq:1}
\end{equation}
rewrite \eqref{eq:1} in terms of the background covariant derivative $D_{\mu} := \partial_{\mu} + ie\mathcal{A}_{\mu}$
\begin{equation}
    \mathcal{L} = -\frac{1}{4} F^{\mu\nu} F_{\mu\nu} + \bar{\psi}(i\gamma^{\mu}\mathcal{D}_{\mu} - m)\psi - e\bar{\psi}\gamma^{\mu}A_{\mu}\psi\label{eq:2}
\end{equation}
This trivial reorganisation of terms implicitly defines the ``Furry picture" expansion of scattering amplitudes. The first two terms in \eqref{eq:2} are still quadratic in the dynamical fields and in this sense represent a “free theory”. 

In IFQED, the Dirac equation with a background field $\mathcal{A}$ is 
\begin{equation}
    [i \slashed{\partial} - e\slashed{\mathcal{A}} - m] \psi(x) = 0\label{eq:DE}
\end{equation}

We first transform \eqref{eq:DE} into a a second order differential equation by multiplying with the adjoint Dirac operator $(i\slashed{\partial}-e\slashed{\mathcal{A}} +m)$, yielding
\begin{equation}
    \left[ (i \partial - e\mathcal{A})^2 - m^2 - \frac{e}{2} \sigma^{\mu\nu} \mathcal{F}_{\mu\nu}(x) \right] \psi(x) = 0\label{eq:DE2}
\end{equation}
with $\sigma^{\mu\nu} = \frac{i}{2}[\gamma^{\mu}, \gamma^{\nu}]$. 
We then make an ansatz for the (positive energy) solution of \eqref{eq:DE2} in the form $\psi(x) = e^{-ipx} \Omega(\phi) u^s_p$ where $p^\mu$ is a constant four-momentum vector, $\Omega$ is a $(4 \times 4)$ Dirac matrix depending only on the laser phase $\phi=k \cdot x$, and $u^s_p$ are the free Dirac spinors for momentum $p$ and spin indices $s=1,2$, fulfilling the free Dirac equation $(\slashed{p}-m)u^s_p=0$ and the orthogonal normalization condition $\bar{u}^s(p) u^{s'}(p)=2m \delta^{s s'}$. 

For the Volkov problem with $k^2=0$, we arrive at a first order ordinary differential equation  for the  matrix function $\Omega$:
\begin{equation}
     2ik\cdot p\frac{d\Omega}{d\phi}=(2ep\cdot \mathcal{A}-e^2\mathcal{A}^2-i\slashed{k}\slashed{A})\Omega
\end{equation}
which can be easily integrated and get
\begin{equation}
    \Omega(\phi)=(1+\frac{e\slashed{k} \slashed{\mathcal{A}}}{2k \cdot p})\exp[-i\int \frac{d\phi}{2k\cdot p} (2ep\cdot \mathcal{A} - e^2\mathcal{A}^2)]
\end{equation}

Introducing the new phase factors and bispinors
\begin{eqnarray}
    S^{(\pm)}&=&p\cdot x +\int \frac{d \phi}{2 p\cdot k}(\pm2e \mathcal{A} \cdot p - e^2\mathcal{A}^2)\\
    &=&q\cdot x \pm e(\frac{a_1 \cdot p}{k\cdot p} \sin{\phi}-\frac{a_2 \cdot p}{k\cdot p} \cos{\phi}),\notag\\
    U^s_p &=& \left( 1 + \frac{e\slashed{k} \slashed{\mathcal{A}}}{2 k \cdot p} \right) u^s_p, 
    V^s_p = \left( 1 - \frac{e\slashed{k} \slashed{\mathcal{A}}}{2 k \cdot p} \right) v^s_p,\notag
\label{eq:UV}
\end{eqnarray}
where $q_\mu=p_\mu + \frac{e^2 a^2}{2 p\cdot k}k_\mu$ is the quasimomentum,
the positive energy Volkov wave function can be writtens as
\begin{equation}
    \psi^{(+)s}_p(x) =U^s_pe^{-iS^{(+)}}
\end{equation}
while the negative energy solutions can be obtained  via the transformation $p\rightarrow -p$:
\begin{equation}
    \psi^{(-)s}_p(x)=\psi^{(+)s}_{-p}(x)=V^s_pe^{iS^{(-)}}
\end{equation}

In the special coordinate frame where $\bf k$ is along the negative $z$-axis, we introduce the special bispinors 
\begin{eqnarray}
    \sqrt{2 p_+}u_p^1 &=&
    \begin{pmatrix}
    m + p_+ \\
    p_1 + i p_2 \\
    -m + p_+ \\
    p_1 + i p_2
    \end{pmatrix}, 
    \sqrt{2 p_+} u_p^2 = 
    \begin{pmatrix}
    -p_1 + i p_2 \\
    m + p_+ \\
    p_1 - i p_2 \\
    m - p_+
    \end{pmatrix}, \notag\\
   \sqrt{2 p_+} v_p^1 &=& 
    \begin{pmatrix}
    m - p_+ \\
    -p_1 - i p_2 \\
    -m - p_+ \\
    -p_1 - i p_2
    \end{pmatrix}, 
    \sqrt{2 p_+}v_p^2 = 
    \begin{pmatrix}
    p_1 - i p_2 \\
    m - p_+ \\
    -p_1 + i p_2 \\
    m + p_+
    \end{pmatrix}
\end{eqnarray}
where $p_\pm=p_0\pm p_3$. Then \eqref{eq:UV} leads to 
\begin{eqnarray}
    U_p^1 = \frac{1}{\sqrt{2 p_+}} 
    \begin{pmatrix}
    m + p_+ \\
    \pi_1^- + i \pi_2^- \\
    -m + p_+ \\
    \pi_1^- + i \pi_2^-
    \end{pmatrix}, 
    U_p^2 = \frac{1}{\sqrt{2 p_+}} 
    \begin{pmatrix}
    -\pi_1^- + i \pi_2^- \\
    m + p_+ \\
    \pi_1^- - i \pi_2^- \\
    m - p_+
    \end{pmatrix},\\
    V_p^1 = \frac{1}{\sqrt{2 p_+}} 
    \begin{pmatrix}
    m - p_+ \\
    -\pi^+_1 - i \pi^+_2 \\
    -m - p_+ \\
    -\pi^+_1 - i \pi^+_2
    \end{pmatrix},
    V_p^2 = \frac{1}{\sqrt{2 p_+}} 
    \begin{pmatrix}
    \pi^+_1 - i \pi^+_2 \\
    m - p_+ \\
    -\pi^+_1 + i \pi^+_2 \\
    m + p_+
    \end{pmatrix}
\end{eqnarray}
where $\pi_\mu^\pm=p_\mu\pm e\mathcal{A}_\mu$.

\subsection{Quantization of Volkov state}

In IFQED, the free fermion field operators are linear combinations of the Volkov wave functions with  the coefficient operators denoting electron annihilation $b^s_{\mathbf{p}}$ and positron creation $d^{s\dagger}_{\mathbf{p}}$:
\begin{align}
    \psi(x) &=\int \frac{d^3p}{(2\pi)^3}\frac{1}{\sqrt{2E_{\bf p}}}\sum_{s=1}^2 \left(b^s_{\mathbf{p}}\psi^{(+)s}_p(x)+d^{s\dagger}_{\mathbf{p}}\psi^{(-)s}_p(x)\right)\\
    \bar\psi(x) &=\int \frac{d^3p'}{(2\pi)^3}\frac{1}{\sqrt{2E_{\bf p'}}}\sum_{s'=1}^2 \left(b^{s'\dagger}_{\mathbf{p'}}\bar\psi^{(+)s'}_{p'}(x)+d^{s'}_{\mathbf{p'}}\bar\psi^{(-)s'}_{p'}(x)\right)
\end{align}
where $\bar\psi^{(\pm)s}_p=\psi^{(\pm)s\dagger}_p\gamma^0$.

The creation and  annihilation operators obey the anticommutation rules
\begin{equation}
    \{d^r_{\mathbf{p}},d^{s\dagger}_{\mathbf{p'}}\}=\{b^r_{\mathbf{p}},b^{s\dagger}_{\mathbf{p'}}\}=(2\pi)^3\delta^{(3)}(\mathbf{p}-\mathbf{p}') \delta^{rs}
\end{equation}
Using the orthogonality and completeness of Volkov wave function~\cite{yakaboylu2015}, we can find that $\psi$ and $\psi^\dagger$ satisfy the equal-time anti-commutation relations
\begin{equation}
    \{ \psi(x), \psi^\dagger (y) \}|_{x_0=y_0} =  \delta^{(3)}(\mathbf{x} - \mathbf{y})
\end{equation}

\subsection{The interaction Hamiltonian}

Unlike the electron case, photons free streaming in the strong laser due to its long mean free path. Thus the field operator for free photon is given by
\begin{equation}
    A(x) = \int \frac{d^3k'}{(2\pi)^3}\frac{1}{\sqrt{2 E_{k'}}}\sum_{r=1}^2 \left( a_{\mathbf{k'}}^r\epsilon^r_{k'} e^{-ik' \cdot x} + a_{\mathbf{k'}}^{r \dagger} \epsilon_{k'}^{r*} e^{ik' \cdot x} \right)
\end{equation}
where $a_{\mathbf{k'}}^r$ are the bosonic annihilation operators that follow the commutation rules
\begin{equation}
    [a_{\mathbf{k'}}^r,a_{\mathbf{k}''}^{r' \dagger}]=(2\pi)^3\delta^{(3)}(\mathbf{k'}-\mathbf{k}'') \delta^{r r'}\label{eq:15}
\end{equation}

The IFQED Hamiltonian describing interactions between photons and fermions is given by
\begin{equation}
    H_{int}=\int d^3 x e\bar{\psi}\gamma^{\mu}A_{\mu}\psi \label{eq:12}
\end{equation}

We introduce the following quadruplets to simplify our notations:
\begin{align}
    \hat{A}^n_{\mathbf{k}'} &= \{ a^1_{\mathbf{k}'}, a^{1\dagger}_{\mathbf{k}'}, a^2_{\mathbf{k}'}, a^{2 \dagger}_{\mathbf{k}'} \},& \varepsilon^n_{k'} &= \{ \varepsilon^1_{k'}, \varepsilon^{1*}_{k'}, \varepsilon^2_{k'}, \varepsilon^{2*}_{k'} \}\\
    \hat{B}^m_{\mathbf{p}} &= \{ b^1_{\mathbf{p}}, d^{1\dagger}_{\mathbf{p}}, b^2_{\mathbf{p}}, d^{2\dagger}_{\mathbf{p}}\},& U^m_p&=\{U^1_p,V^1_p,U^2_p,V^2_p\}\\
    \hat{B}^{m'\dagger}_{\mathbf{p}'}&=\{b^{1\dagger}_{\mathbf{p}'},d^1_{\mathbf{p}'},b^{2\dagger}_{\mathbf{p}'},d^2_{\mathbf{p}'}\},& \bar{U}^{m'}_{p'}&=\{\bar{U}^1_{p'},\bar{V}^1_{p'},\bar{U}^2_{p'},\bar{V}^2_{p'}\}
\end{align}
with $n,m,m'=1,2,3,4$.

Using these new quadruplets, the Hamiltonian can be rewritten as
\begin{equation}
    H_{int}=e \int d^3 x 
 \int \frac{d^3k'\, d^3p' \, d^3p}{(2\pi)^9} 
\sum_{n,m',m} \hat{A}^n_{\mathbf{k}'} \hat{B}^{m'\dagger}_{\mathbf{p}'} \hat{B}^m_{\mathbf{p}} C^{nm'm} e^{i \Phi^{nm'm}}\label{eq:H1}
\end{equation}
where the coefficient ``tensor" $C^{nm'm}$ is defined as $C^{nm'm}\equiv\bar{U}^{m'}_{p'} \slashed{\varepsilon}^n_{k'} U^m_p/ \sqrt{8E_{k'}E_{p}E_{p'}}$. For an arbitrary polarization vector decomposed as $\varepsilon^n_{k'}=\epsilon_L\varepsilon^L+\epsilon_R\varepsilon^R+\epsilon_k\varepsilon^k$, with $\varepsilon^{R/L}=(0,1,\pm i,0)/\sqrt{2}$, and $\varepsilon_k=(0,0,0,1)$, the coefficient ``tensor" can also be expressed by 
\begin{equation}
(C^n)^{m'm}=\epsilon_L(C^L)^{m'm}+\epsilon_R(C^R)^{m'm}+\epsilon_k(C^k)^{m'm}
\end{equation}
And the phase factors $\Phi^{nm'm}$ are given by
\begin{equation}
    \Phi^{nm'm} = (-1)^n k'-(-1)^{m'}q'+(-1)^mq+\alpha_1 \sin \phi - \alpha_2 \cos \phi
\end{equation}
where $\alpha_i = e a_i\cdot(\frac{p'}{k\cdot p'}-\frac{p}{k\cdot p}), \, i=1,2$.

We expand phase functions in Fourier series and denote the expansion coefficients by Bessel functions. As a result, the coefficients are expanded as
\begin{eqnarray}
    C^{nm'm} e^{i \Phi^{nm'm}}=
    \sum_{l=-\infty}^{\infty}\sum_{w=-1}^{1} C_w^{nm'm} D_w^l e^{ixQ}\label{eq:coef}
\end{eqnarray}
where $Q \equiv \{Q^0, \mathbf{Q}\} = l k + (-1)^n k' - (-1)^{m'} q' + (-1)^m q$.

Substituting \eqref{eq:coef} into \eqref{eq:H1} and integrating the spatial part, we arrive at the final form of the Hamiltonian
\begin{align}
    H_{int}= &e \int \frac{ d^3 k' \, d^3 p' \, d^3 p}{(2\pi)^9}
    \sum_{nm'm}\hat{A}^n_{\mathbf{k}'} \hat{B}^{m'\dagger}_{\mathbf{p}'} \hat{B}^m_{\mathbf{p}} \sum_{l,w}C_w^{nm'm} D_w^l e^{iQ^{0}t}\,
	\delta^{(3)}(\mathbf{Q}).
\end{align}

Assuming a finite cubic box of volume $V=L^3$, make the replacement 
\begin{align}
    &p_i\rightarrow \frac{2\pi \tilde p_i}{L},\quad \int\frac{d^3p}{(2\pi)^3}\rightarrow \frac{1}{V}\sum_{\mathbf{\tilde p}}, \\ \nonumber
    &b_\mathbf{p}^s\rightarrow \sqrt{V}\tilde{b}_{\mathbf{\tilde p}}^s, \quad d_\mathbf{p}^s\rightarrow \sqrt{V}\tilde{d}_{\mathbf{\tilde p}}^s
\end{align}

We omit the tildes from now on when there is no ambiguities. The Hamiltonian now reads
\begin{align}\label{discretized H}
    H_{int}\rightarrow \frac{e}{\sqrt{V}}
    &\sum_{k'p'p,nm'm}\hat{A}^n_{\mathbf{k}'} \hat{B}^{m'\dagger}_{\mathbf{p}'} \hat{B}^m_{\mathbf{p}} \sum_{l,w} C_w^{nm'm} D_w^l e^{i Q^{0} t} \delta^{(3)}_{\mathbf{Q}}
\end{align}

\section{spin-phonon Hamiltonian}
\label{sec:analog gate}

In trapped-ion systems, we can couple the phonon mode of ions to its internal spin mode, which is derived from the ion-laser interaction Hamiltonian, to construct an analog spin-phonon gate. 

The ion-laser Hamiltonian is given by the interaction between the ion's internal spin (for $j$-th ion) and electromagnetic field
\begin{equation}
    H'_{\text{ion-laser}} = \frac{1}{2} \hbar \Omega_j \, (\sigma^{+}_j + \sigma^{-}_j) \left[ e^{i(k \hat{z} - \omega_j^L t + \phi_j)} + e^{-i(k \hat{z} - \omega_j^L t + \phi_j)} \right],
\end{equation}
where we assume laser as a classical source of electromagnetic waves. The coupling strength $\Omega_j$ is also known as the Rabi frequency and the $\sigma^{+}_j / \sigma^{-}_j$ are spin raising/lowering operators. The wave-vector, frequency, and phase of the electromagnetic plane wave are $k$, $\omega_{j}^L$ and $\phi_j$, respectively. Actually, we usually two tones of off-resonant laser to drive Raman transitions between the spin levels in experiment. In this situation, $\omega_{j}^L$ represents the frequency diﬀerence (beat-note) of laser beams.

The position operator ($\hat{z}$) can be rewritten in terms of the harmonic oscillator creation and annihilation operators $a_{\mathrm{k}}^{r\dagger}$, $a^r_{\mathrm{k}}$. 
\begin{equation}
    k \cdot \hat{z} = k z_0 cos\theta' (a^r_{\mathrm{k}} + a_{\mathrm{k}}^{r\dagger}) = \eta_{k,j} (a^r_{\mathrm{k}} + a_{\mathrm{k}}^{r\dagger})
\end{equation}
where $\eta_{k,j}$ is Lamb-Dicke parameter, which measures the degree of coupling between the vibration modes of ions and electromagnetic fields. $z_0 = \sqrt{\hbar /2m \omega_k}$ is the displacement amplitude of ions and $\theta'$ is the angle between the wavevector(usually assume $\theta'=0$) and the direction of vibration mode. 

Moving into the interaction picture with free Hamiltonian $H_0 = \frac{1}{2}\hbar \omega_0 \sigma^z_j + \hbar \omega_k (a_{\mathrm{k}}^{r\dagger}a^r_{\mathrm{k}}+\frac{1}{2})$ and making the rotating-wave approximation (RWA) to neglect high frequency oscillating terms, we get the interaction Hamiltonian
\begin{align}
    H'_{\text{ion-laser}} = \frac{1}{2} \hbar \Omega_j \, \sigma^{+}_j  e^{i \eta_{k,j}(a^r_{\mathrm{k}}e^{-i\omega_{\mathrm{k}} t} + a_{\mathrm{k}}^{r\dagger} e^{i\omega_{\mathrm{k}} t})} e^{-i(\omega_j^L - \omega_0) t + i\phi_j} + \mathrm{h.c.} \, ,
\end{align}
where $\omega_{\mathrm{k}}$  and $\omega_0$, correspond respectively to the vibrational mode frequency, and the energy difference between the two internal levels of the ion.

Within Lamb-Dicke limit, we can make the approximation
\begin{equation}
    e^{i \eta_{k,j}(a^r_{\mathrm{k}}e^{-i\omega_{\mathrm{k}} t} + a_{\mathrm{k}}^{r\dagger} e^{i\omega_{\mathrm{k}} t})} \sim 1 + i \eta_{k,j}(a^r_{\mathrm{k}}e^{-i\omega_{\mathrm{k}} t} + a_{\mathrm{k}}^{r\dagger} e^{i\omega_{\mathrm{k}} t}).
\end{equation} 
Hence, the Hamiltonian is splited into two parts
\begin{align}
    H'_{\text{ion-laser}} = &\frac{1}{2} \hbar \Omega_j \, \sigma^{+}_j e^{-i(\omega_j^L - \omega_0) t + i\phi_j} + \mathrm{h.c.} \nonumber \\
    & + i \frac{\hbar}{2} \eta_{\mathrm{k},j} \Omega_{j} \sigma_j^+ \left( a^r_{\mathrm{k}} e^{-i\omega_{\mathrm{k}} t} + a_{\mathrm{k}}^{r\dagger} e^{i\omega_{\mathrm{k}} t} \right) e^{-i(\omega_j^L-\omega_0) t}e^{i \phi_j} + \mathrm{h.c.}
\end{align}
The fisrt part is called carrier transition. We can realize the transition between the spin states of the ion and do not change the vibration mode by setting the laser detuning $\delta = \omega_{j}^L-\omega_0 = 0$. The second part describes the coupling of spin to phonon in trapped ion. We define it as $H_{\mathrm{k},j}^{\sigma a}$, which makes it convenient to discuss this part separately. By setting the laser detuning $\delta = \omega_{j}^L-\omega_0 = \pm \omega_{\mathrm{k}}$, we can realize blue sideband(BSB) and red sideband(RSB) transitions, respectively.

To realize our analog spin-phonon gate, we need to apply red and blue sideband simultaneously, whcih means that the noncopropagating laser beams have bichromatic beat notes at frequencies $\omega_{j}^L = \omega_0 \pm \omega_{\mathrm{k}}$ symmetrically. Thus, the coupling Hamiltonian $H_{\mathrm{k},j}^{\sigma a}$ can be written as(set $\hbar=1$)
\begin{align}\label{eq:blue-and-red}
    H_{\mathrm{k},j}^{\sigma a} = &i \frac{\eta_{\mathrm{k},j} \Omega_{j}}{2} \left( a^r_{\mathrm{k}} e^{-i(\omega_j^L -\omega_0 +\omega_{\mathrm{k}}) t} e^{i \phi_j^{rsb}}  \sigma_j^+ -  a_{\mathrm{k}}^{r\dagger} e^{i(\omega_j^L -\omega_0 +\omega_{\mathrm{k}})t} e^{- i \phi_j^{rsb}}  \sigma_j^- \right)
    \nonumber \\
    + & i \frac{\eta_{\mathrm{k},j} \Omega_{j}}{2} \left( a_{\mathrm{k}}^{r\dagger} e^{-i(\omega_j^L -\omega_0 -\omega_{\mathrm{k}})t} e^{i \phi_j^{bsb}} \sigma_j^+ - a^r_{\mathrm{k}} e^{i(\omega_j^L -\omega_0 - \omega_{\mathrm{k}}) t} e^{-i\phi_j^{bsb}} \sigma_j^- \right)
\end{align}
The first line represents the red sideband Hamiltonian and the second line represents the blue sideband Hamiltonian. We define spin and vibration phases as
\begin{align}
    \phi_{\mathrm{k},j}^s = \frac{1}{2}(\phi_j^{rsb} + \phi_j^{bsb}) \nonumber \\
    \phi_{\mathrm{k},j}^v = \frac{1}{2}(\phi_j^{rsb} - \phi_j^{bsb}).
\end{align}
The coupling Hamiltonian $H_{\mathrm{k},j}^{\sigma a}$ will become the form of spin component times vibration component
\begin{align}\label{eq:spin-times-vibration-component}
    H_{\mathrm{k},j}^{\sigma a} = &i \frac{\eta_{\mathrm{k},j} \Omega_{j}}{2} \sigma_j^+ e^{i \phi_{\mathrm{k},j}^s} 
    \left( a^r_{\mathrm{k}} e^{-i(\omega_j^L -\omega_0 +\omega_{\mathrm{k}}) t} e^{i \phi_{\mathrm{k},j}^v} + a_{\mathrm{k}}^{r\dagger} e^{-i(\omega_j^L -\omega_0 -\omega_{\mathrm{k}})t} e^{-i \phi_{\mathrm{k},j}^v} \right) 
    \nonumber \\
    & - i \frac{\eta_{\mathrm{k},j} \Omega_{j}}{2} \sigma_j^- e^{- i \phi_{\mathrm{k},j}^s} 
    \left( a^r_{\mathrm{k}} e^{i(\omega_j^L -\omega_0 - \omega_{\mathrm{k}}) t} e^{i \phi_{\mathrm{k},j}^v} + a_{\mathrm{k}}^{r\dagger} e^{i(\omega_j^L -\omega_0 + \omega_{\mathrm{k}})t} e^{-i \phi_{\mathrm{k},j}^v} \right) 
\end{align}
If we require that $\delta - \omega_{\mathrm{k}} = \epsilon$ and $-\epsilon = \omega_{\mathrm{k}} + \delta$, which means that the red and blue sidebands are driven at a detuning with equal magnitude but opposite sign, simultaneously. 
The \eq{spin-times-vibration-component} becomes
\begin{align}
    H_{\mathrm{k},j}^{\sigma a} = &i \frac{\eta_{\mathrm{k},j} \Omega_{j}}{2} \left(\sigma_j^+ e^{i \phi_{\mathrm{k},j}^s} -  \sigma_j^- e^{- i \phi_{\mathrm{k},j}^s} \right) 
    \left( a^r_{\mathrm{k}} e^{i \epsilon t} e^{i \phi_{\mathrm{k},j}^v} + a_{\mathrm{k}}^{r\dagger} e^{-i \epsilon t} e^{-i \phi_{\mathrm{k},j}^v} \right) 
\end{align}
which is the Mølmer-Sørensen Hamiltonian. This interaction provides the common physical basis for both the CNOT gate and the analog gate implemented in our scheme.

We can choose special spin phase $\phi_{\mathrm{k},j}^s=\pi$ to make the Hamiltonian be proportional to $\sigma_y$ and the detuning $\epsilon=0$. With these setting, the Hamiltonian in \eq{blue-and-red} becomes~\cite{PhysRevResearch.3.043072} \cite{monroe2021programmable}
\begin{equation}
    H_{\mathrm{k},j}^{\sigma a} (\phi_{\mathrm{k},j}) = \frac{\eta_{\mathrm{k},j} \Omega_{j}}{2}\left( a^r_{\mathrm{k}} e^{i\phi_{\mathrm{k},j}^v} + a_{\mathrm{k}}^{r\dagger} e^{-i\phi_{\mathrm{k},j}^v} \right) \sigma_j^y
\end{equation}
If we instead choose $\phi_{\mathrm{k},j}^s=\pi/2$, we would obtain an analog gate that couples phonon modes to $\sigma_j^x$.

The analog spin-phonon gate $R_{k,j}^{\sigma a}(\theta_{k,j}, \phi_{k,j}^v)$ can be obtained by applying $H_{k,j}^{\sigma a} (\phi_{k,j}^v)$ for time $\tau_{gate}$ with particular phase $\phi_{k,j}^v$ :
\begin{align}
    R_{k,j}^{\sigma a}(\theta_{k,j}, 0) = e^{-i\theta_{k,j}(a^r_k + a_k^{r\dagger})\sigma_j^y}\\
    R_{k,j}^{\sigma a}(\theta_{k,j}, \frac{\pi}{2}) = e^{\theta_{k,j}(a^r_k - a_k^{r\dagger})\sigma_j^y}
\end{align}
where $\theta_{k,j} = \frac{\eta_{\mathrm{k},j} \Omega_{j}}{2} \times \tau_{gate}$. We can choose appropriate laser parameter to make $\theta_{k,j}$ equal to $\theta$ mentioned in \eq{uhij}, the product of the evolution time and the coupling strength in the Hamiltonian i.e. $\theta_{k,j} \sim \theta = c_{ijk} t$.

In our simulation, we consider two-qubit MS gates are performed with two $^{40}$Ca$^+$ ions, where the 729 nm laser is applied to drive the transition between state $\ket{0}\equiv\ket{S_{1/2}, m = -1/2}$ and state $\ket{1}\equiv\ket{D_{5/2}, m = -1/2}$. If we choose radial COM mode with frequency $\omega_{x}=2\pi\times1.18$ MHz to couple two ions. The Lamb-Dicke parameter can be calculated to be 0.08918($\eta \ll 1$).

\section{Multi-mode configuration}
\label{sec:multi-mode}

\subsection{truncate $l$ to 2}

As mentioned, the effect of the background field enters through the exchange of an effective harmonic number $l$. We have simulated the situation of etaining only the $l=1$ harmonic contribution. To show the scalability of our framework, we should consider more background virtual photons participating in the interaction which will produce more different momentum modes, e.g. the situation of effective harmonic number $l$ is truncated to 2. 

The interaction Hamiltonian of effective harmonic number $l$ is truncated to 2 can be written by \eq{H_BW}
\begin{align}\label{eq:Hamiltonian l=2}
    H_{int,l=2}=&\frac{e}{2\sqrt{V k'^0p_1^0p_1'^0p_{1+} p'_{1+}}} \left[ p'_{1+} (p_{1R} D_0^1-eaD^1_{-1})a_{\mathbf{k'}}^{1 \dagger}d^1_{\mathbf{p_1'}} b^1_{\mathbf{p_1}} + p_{1+}( p'_{1R} D_0^1+eaD^1_{-1})a_{\mathbf{k'}}^{1 \dagger}d^2_{\mathbf{p_1'}} b^2_{\mathbf{p_1}}+\right. \nonumber\\
    &p_{1+}(p'_{1L} D_0^1+ ea D^1_1)a_{\mathbf{k'}}^{2 \dagger} d^1_{\mathbf{p_1'}} b^1_{\mathbf{p_1}} +p'_{1+}(p_{1L}D_0^1- ea D^1_1)a_{\mathbf{k'}}^{2 \dagger} d^2_{\mathbf{p_1'}} b^2_{\mathbf{p_1}} \left.+ m(p_{1+} + p'_{1+})D_0^1(a_{\mathbf{k'}}^{1 \dagger}d^1_{\mathbf{p_1'}} b^2_{\mathbf{p_1}} - a_{\mathbf{k'}}^{2 \dagger} d^2_{\mathbf{p_1'}} b^1_{\mathbf{p_1}})\right]\nonumber\\
    &+\mathrm{h.c.} \quad (l=1)  \nonumber\\
    &+ \frac{e}{2\sqrt{V k'^0p_2^0p_2'^0p_{2+} p'_{2+}}}\times 
    \left[ p'_{2+} (p_{2R} D_0^2-eaD^2_{-1})a_{\mathbf{k'}}^{1 \dagger}d^1_{\mathbf{p_2'}} b^1_{\mathbf{p_2}} + p_{2+}( p'_{2R} D_0^2+eaD^2_{-1})a_{\mathbf{k'}}^{1 \dagger}d^2_{\mathbf{p_2'}} b^2_{\mathbf{p_2}}  \right.\nonumber\\
    &+p_{2+}(p'_{2L} D_0^2+ ea D^2_1)a_{\mathbf{k'}}^{2 \dagger} d^1_{\mathbf{p_2'}} b^1_{\mathbf{p_2}}+p'_{2+}(p_{2L}D_0^2- ea D^2_1)a_{\mathbf{k'}}^{2 \dagger} d^2_{\mathbf{p_2'}} b^2_{\mathbf{p_2}}
    \left.+ m(p_{2+} + p'_{2+})D_0^2(a_{\mathbf{k'}}^{1 \dagger}d^1_{\mathbf{p_2'}} b^2_{\mathbf{p_2}} - a_{\mathbf{k'}}^{2 \dagger} d^2_{\mathbf{p_2'}} b^1_{\mathbf{p_2}})\right]\nonumber\\
    &+\mathrm{h.c.} \quad (l=2)
\end{align}
where subscript represents different harmonic contribution $l$. Because different numbers of background virtual photons $l$ participating in the interaction will produce different momentums of fermions and antifermions. 

The parameters of background laser field and incident photon are same as $l=1$: $\xi = 1$, $\omega = 1.55~\mathrm{eV}$ and $\omega' = 1~\mathrm{TeV}$. The outgoing fermion and antifermion have opposite transverse momentums due to quasi-momentum conservation $q_l+q_l'=lk+k'$ for the head-on high-energy photon and laser collision. And we ssume that fermion and antifermion have same longitudinal momentum. For $l=1$, momentum conservation $q_1+q'_1=k+k'$ yields
\begin{align}
	q_1^0 &= {q'}_1^0 = \frac{\omega'+\omega}{2} \simeq 0.5~\mathrm{TeV}, \\
	q_1^3 &= {q'}_1^3 = \frac{\omega'-\omega}{2} \simeq 0.5~\mathrm{TeV}, \\
	|\vec q_1^{\perp}|
	&= |\vec {q'}_1^{\perp}|
	= \sqrt{\omega\omega'-m_*^2}
	= 1.014~\mathrm{MeV},
	\label{eq:l1_num}
\end{align}
For $l=2$, momentum conservation $q_2+q'_2=2k+k'$ yields
\begin{align}
	q_2^0 &= {q'}_2^0 = \frac{\omega'+2\omega}{2} \simeq 0.5~\mathrm{TeV}, \\
	q_2^3 &= {q'}_2^3 = \frac{\omega'-2\omega}{2} \simeq 0.5~\mathrm{TeV}, \\
	|\vec q_2^{\perp}|
	&= |\vec {q'}_2^{\perp}|
	= \sqrt{2\omega\omega'-m_*^2}
	= 1.606~\mathrm{MeV}.
	\label{eq:l2_num}
\end{align}
The characteristic energy scale of the process is of order TeV, 
and different momentum modes are mainly distinguished by transverse momentums at the MeV scale. Thus, we take the discretize momentum spacing $\Delta p_n = 0.03~\mathrm{MeV}$.
This corresponds to $L=\frac{2\pi}{\Delta p_n}=4.12\times10^{-11}\,\mathrm{m}$ and a lattice volume $V=L^3$. Hence, the overall prefactor of the interaction Hamiltonian is then estimated as
\begin{align}
	\frac{e}{2\sqrt{V k'^0p_1^0p_1'^0p_{1+} p'_{1+}}}\sim 10^{-25}~\mathrm{eV}^{-1}, \quad
    \frac{e}{2\sqrt{V k'^0p_2^0p_2'^0p_{2+} p'_{2+}}}
	\sim 10^{-25}~\mathrm{eV}^{-1}.
\end{align}

Under these assumptions, the interaction Hamiltonian \eq{Hamiltonian l=2} reduces to twelve dominant terms

\begin{align}
    H_{\mathrm{int},l_{max}=2}=&(-1.35a_{\mathbf{k'}}^{L \dagger}d^1_{\mathbf{p_1'}} b^1_{\mathbf{p_1}}
	+3.23 a_{\mathbf{k'}}^{L \dagger}d^1_{\mathbf{p_1'}} b^2_{\mathbf{p_1}}
	+1.35 a_{\mathbf{k'}}^{L \dagger}d^2_{\mathbf{p_1'}} b^2_{\mathbf{p_1}} \nonumber\\
    &-2.93a_{\mathbf{k'}}^{R \dagger} d^1_{\mathbf{p_1'}} b^1_{\mathbf{p_1}}
	-3.23a_{\mathbf{k'}}^{R \dagger}d^2_{\mathbf{p_1'}} b^1_{\mathbf{p_1}}+2.93a_{\mathbf{k'}}^{R \dagger} d^2_{\mathbf{p_1'}} b^2_{\mathbf{p_1}})\times 10^{-8} \mathrm{eV} + \mathrm{h.c.} \nonumber\\
    &+(-0.30a_{\mathbf{k'}}^{L \dagger}d^1_{\mathbf{p_2'}} b^1_{\mathbf{p_2}}
	+1.30 a_{\mathbf{k'}}^{L \dagger}d^1_{\mathbf{p_2'}} b^2_{\mathbf{p_2}}
	+0.30 a_{\mathbf{k'}}^{L \dagger}d^2_{\mathbf{p_2'}} b^2_{\mathbf{p_2}}\nonumber\\
	&-1.93a_{\mathbf{k'}}^{R \dagger} d^1_{\mathbf{p_2'}} b^1_{\mathbf{p_2}}
	-1.30a_{\mathbf{k'}}^{R \dagger}d^2_{\mathbf{p_2'}} b^1_{\mathbf{p_2}}+1.93a_{\mathbf{k'}}^{R \dagger} d^2_{\mathbf{p_2'}} b^2_{\mathbf{p_2}})\times 10^{-8} \mathrm{eV} + \mathrm{h.c.}
\end{align}
The coefficient of interaction Hamiltonian is related Bessel function of the first kind. The order of Bessel function and its variable are both related to $l$. The function will rapidly decay as $l$ increases, and the function decays to 0 when $l=6$. Thus, the coefficients of $l=1$ sectors are larger than $l=2$ sectors.

We take the initial state to be a left-polarized photon $|\gamma^L\rangle=a_{\mathbf{k}'}^{L\dagger}|\Omega\rangle$,
where the vacuum state $|\Omega\rangle$ satisfies $\hat{a}^n_{\mathbf{k}'}|\Omega\rangle= \hat{b}^{m'}_{\mathbf{p}'}|\Omega\rangle= \hat{d}^m_{\mathbf{p}}|\Omega\rangle=0$. We then compute the probabilities for the channels $\gamma^L \to \gamma^L$, $\gamma^L \to \gamma^R$, and $\gamma^L \to f_{p_l}^s\bar{f}_{p'_l}^{s'}$
where $|f_{p_l}^s\bar{f}_{p'_l}^{s'}\rangle = b^{s\dagger}_{\mathbf{p}_l}d^{s'\dagger}_{\mathbf{p'}_l}|\Omega\rangle$.

These probabilities are obtained by direct diagonalization of $H_{\mathrm{int},l_{max}=2}$, and the resulting dynamics are shown in \fig{evo}($\bm{b}$). Solid lines correspond to fermion-antifermion pair satisfied momentum conservation $q_1+q'_1=k+k'$, and dash-dotted lines correspond to fermion-antifermion pair satisfied momentum conservation $q_2+q'_2=2k+k'$. Orange line represents the decay of left-polarized photon, and purple line means that there is a certain probability that left-polarized photon will transform into right-polarized photon when entering the strong background field. We can capture significant polarization dependence of the nonlinear Breit--Wheeler process and the probability of more background virtual photons participating the interaction is depressed. In \fig{evo}($\bm{c}$), we compare the situation of $l$ truncated to 1 and 2. For $l$ truncated to 2, the decay rate of incident left-polarized photon is larger than only $l=1$ harmonic contribution. It is natural, because more background virtual photons participating in the interaction will increase the decay channels of incident left-polarized photon.

\subsection{Quantum circuit construction}

We order the fermionic modes sequentially along a one-dimensional spin chain, with the four internal states associated with each momentum mode occupying four consecutive qubits. For the case where $l$ is truncated to 2, there are at least two set momentum modes, that is, we need to extend the spin chain to 8 ions. And the fermionic degrees of freedom are represented by ion spins via a Jordan--Wigner transformation as followed
\begin{align}
    b_{p_1}^{1\dagger} &= \sigma^+ (\otimes \mathcal{I})^7   \qquad
    d_{p_1'}^{1\dagger} = \sigma^z \otimes \sigma^+ (\otimes \mathcal{I})^6 \qquad
    b_{p_1}^{2\dagger} = (\sigma^z \otimes)^{2} \sigma^+ (\otimes \mathcal{I})^5 \qquad
    d_{p_1'}^{2\dagger} = (\sigma^z \otimes)^{3} \sigma^+ (\otimes \mathcal{I})^4 \nonumber \\
    b_{p_2}^{1\dagger} &= (\sigma^z \otimes)^{4} \sigma^+ (\otimes \mathcal{I})^3 \qquad
    d_{p_2'}^{1\dagger} = (\sigma^z \otimes)^{5} \sigma^+ (\otimes \mathcal{I})^2 \qquad
    b_{p_2}^{2\dagger} = (\sigma^z \otimes)^{6} \sigma^+ (\otimes \mathcal{I})^1 \qquad
    d_{p_2'}^{2\dagger} = (\sigma^z \otimes)^{7} \sigma^+. 
\end{align}
After that, the time evolution under the full Hamiltonian can be constructed via a Trotter--Suzuki decomposition. For each evolution time $t$, we can achieve the error $\varepsilon$ between the quantum circuit $\mathcal{U}(t)$ and the unitary evolution $U$ to 0.01 when Trotter steps is set to $N_t \sim \mathcal{O}\!\left(t^2 N_p^2 C/\epsilon\right)$, where $N_p$ is the total number of momentum modes and $C \sim 0.01$ denotes a typical magnitude of $\|[H_{ij},H_{i'j'}]\|$ among noncommuting terms. In the actual construction of quantum circuits, we used $\lceil (4t)^2 \rceil$-step Trotter at dynamic time $t$ which result $\lceil (4t)^2 \rceil \times 108$ CNOT gates and $\lceil (4t)^2 \rceil \times 48$ Analog gates.

\section{Error Mitigation}
\label{sec:error mitigation}

Quantum devices inevitably suffer from unwanted interactions with the environments, which manifests as noise that degrades the fidelity of quantum states and quantum operations. Thus, the expectation value of an observable $\hat{O}$ derived from the noisy state deviates from its ideal value. We denote this as $O (\lambda) = \text{tr}(\hat{O} \rho(t, \lambda))$, where $\lambda$ represents the effective noise strength and $\rho(t, \lambda)$ as the density matrix for the noisy state at the evolution time $t$. Our objective is to employ error mitigation techniques to infer the expectation value $O(\lambda)$ in the limit of $\lambda \to 0$, recovering the results obtained from noiseless quantum devices.

\subsection{Noise Model}
\label{sec:noise model}

In trapped-ion quantum platforms, quantum coherence is affected by several distinct noise channels arising from environmental coupling and technical imperfections. In the present work, we focus on two main classes of errors. The first involves energy-exchange processes, including the anomalous heating of phonon modes caused by fluctuating ambient electric fields at the ion position, and the decay of qubits from the excited state to the ground state which is of a $T_1$-type error. The second is dephasing (a $T_2$-type error), which erodes the phase information of superposition states without energy loss. For phonon modes, dephasing is often induced by trap-frequency instability or fluctuations in the trapping potentials. For spin modes, dephasing typically stems from magnetic field noise and laser instabilities, such as phase noise and frequency jitter. 

In our hybrid analog-digital circuit, we use a specific phonon mode (the stretch mode) to represent the incident photon and store its dynamical information as the storage phonon mode. Each CNOT operation is implemented based on the native Mølmer-Sørensen gate in trapped ion devices, which requires one additional phonon mode as an auxiliary phonon mode. For the noise channels, since the storage phonon mode cannot be recooled during the circuit evolution, it is continuously affected by environment noise. In contrast, the noise channel for the auxiliary phonon mode is active only during the operation of the CNOT gate.

In the noisy circuit simulation, we employ the Lindblad equation to incorporate the effects of noises. For convenience, we consider the Lindblad equation for the density matrix in terms of the duration time $\tau$ of a quantum circuit, which is $\rho(\tau, \lambda)$. To manifest the effects of the noise on the quantum states, we also present the noise channels considered as Kraus operator, i.e. $\mathcal{E}(\rho)=\sum_k E_k \rho E_k^{\dagger}$, where $\sum_k E_k E_k^{\dagger} = \mathbb{I}$. 

\subsubsection{Energy-exchange channel}

\paragraph{Phonon mode}
The energy-exchange channel for the phonon mode induces a stochastic increase in the phonon occupancy and is modeled as the heating rate channel. 

The origin of this noise can be traced back to voltage fluctuations on the trap electrodes, which generate a stochastic electric field coupled to the ion displacement. The interaction between this noisy electric field and the ions can be effectively modeled by a classical field \(E_{\rm cl}\) coupled to the phonon mode through
$V \sim -q(a+a^\dagger)E_{\rm cl}$
where \(q\) is the ion charge~\cite{PhysRevA.62.053807}. Since the classical field couples symmetrically to the phonon creation and annihilation operators, it induces heating and cooling processes at the same rate \(\Gamma_{\rm heat}\), which is equivalent to trapped ion system coupled to an infinite high-temperature bath. As a result, the time evolution of the phonon density matrix can be described by the following Lindblad equation for this heating rate channel
\begin{align}\label{eq:lindblad-heating-rate}
    \frac{d\rho_p}{d\tau}
    =
    \Gamma_{\rm heat}\,\mathcal{L}[a]\rho_p
    +
    \Gamma_{\rm heat}\,\mathcal{L}[a^\dagger]\rho_p,
    \qquad
    {\rm with\,\,\,}\mathcal{L}[L]\rho_p
    =
    L\rho_p L^\dagger
    -
    \frac{1}{2}\{L^\dagger L,\rho_p\}.
\end{align}
In the noisy circuit simulation, we assume that $\Gamma_{\rm heat} = 3\, {\rm phonons} /s$, meaning that heating the phonon mode from the ground state to the first excited state takes $1/3$ seconds.

To translate this heating rate channel to Kraus operator, we first write the density matrix of phonon mode in coherent state representation(P-representation)~\cite{breuer2002opensys}
\begin{equation}
    \rho_p(\tau) = \int d^2 \alpha P(\alpha, \alpha^*,\tau) |\alpha\rangle \langle \alpha |,
\end{equation}
where $P(\alpha, \alpha^*,\tau)$ is weight function. In P-representation, the solution of \eq{lindblad-heating-rate} can be obtained
\begin{equation}\label{eq:ds-heating-rate}
    \rho_p(\tau) = \int d^2 \alpha \frac{1}{\pi \sigma^2(\tau)} \exp{(-\frac{|\alpha|^2}{\sigma^2(\tau)})} D(\alpha) \rho_p(0) D^{\dagger}(\alpha),
\end{equation}
where $\sigma^2(\tau) = \Gamma_{\rm heat} \tau$ increases linearly at the rate of $\Gamma_{\rm heat}$ over time and $\hat{D}(\alpha) = \exp(\alpha a^\dagger - \alpha^* a)$ is the displacement operator. If the initial state is vacuum state $\rho_{p0}=|0\rangle \langle0|$, the width $\sigma^2(\tau)$ can  interpreted as the average number of phonon in the system at duration time $\tau$. 

According to \eq{ds-heating-rate}, the Kraus operators of this heating rate channel in the coherent state basis can be defined as
\begin{equation}
    \hat{E}(\alpha) = \sqrt{\frac{1}{\pi \sigma^2(\tau)} \exp\left(-\frac{|\alpha|^2}{\sigma^2(\tau)}\right)} \hat{D}(\alpha).
\end{equation}
The completeness relation is satisfied by $\int d^2\alpha \hat{E}^\dagger(\alpha) \hat{E}(\alpha) = \mathbb{I}$. This representation provides a clear physical intuition: the heating process is equivalent to a series of stochastic phase-space displacement $\alpha$ generated by $\hat{D}(\alpha)$. These displacements are not deterministic but follow a two-dimensional Gaussian distribution centered at the origin of phase space, with a time-dependent variance $\sigma^2(\tau)$. In the trapped-ion system, both the analog gates (\(\mathbf{A}\)) and the CNOT gate involve displacement operations. Consequently, heating noises during these gate operations can directly distort the phonon phase-space trajectories, resulting in residual spin-phonon entanglement and a corresponding degradation of the gate fidelity.

We can also write the Kraus operator $\hat{E}_k$ of heating rate channel in the Fock state basis. For the initial state density matrix $\rho_{p0}=|n\rangle \langle n|$, $\hat{E}_k$ can be written as
\begin{align}
    \hat{E_k}=\sqrt{\frac{(\sigma^2)^k}{(1+\sigma^2)^{k+1}}\sum_{m=0}^n \binom{n}{m} \binom{n+k}{n-m}\frac{(\sigma^2)^{2m}}{(\sigma^2 + 1)^{2n}}} |n +k\rangle \langle n|, \\
    \hat{E}_{-k}=\sqrt{\frac{(\sigma^2)^k}{(1+\sigma^2)^{k+1}}\sum_{m=0}^{n-k} \binom{n-k}{m} \binom{n}{k+m} \frac{(\sigma^2)^{2m}}{(\sigma^2 + 1)^{2(n-k)}}} |n -k\rangle \langle n|,
\end{align}
where $k\ge 0$ represents the order of the transition. The completeness relation is satisfied by $\sum_k \hat{E}^\dagger_k \hat{E}_k= \mathbb{I}$. The heating rate channel can then be written as $\rho_p=\mathcal{E}(\rho_{p0})=\sum_{k=0}^{\infty} \hat{E}_k \rho_{p0} \hat{E}_k^{\dagger} + \sum_{k=1}^{n} \hat{E}_{-k} \rho_{p0} \hat{E}_{-k}^{\dagger}$.

\paragraph{Spin mode}
The spin degree of freedom may also undergo an energy exchange process with environment ($T_1$-type error), corresponding to the decay of the qubit from the excited state to the ground state.
For optical qubits in trapped-ion systems, this process is primarily set by the finite lifetime of the metastable excited state. The corresponding relaxation time $T_1$ is typically on the order of seconds~\cite{bruzewicz2019trapped}, whereas the gate operations considered in our simulations take place on the microsecond timescale. Since the gate time is much shorter than the relaxation time, the spin-state decay is neglected in the noisy circuit simulation.

\subsubsection{Dephasing Error}

\paragraph{Phonon mode}

Besides phonon heating, phonon modes are also subject to dephasing arising from trap-frequency instability and fluctuations of the trapping potential. The corresponding dephasing dynamics of the phonon mode is described by the following Lindblad equation
\begin{equation}\label{eq:lindblad-phonon-dephasing}
    \frac{d\rho_p}{d\tau} = \Gamma_{dp} \mathcal{L}[a^{\dagger}a]\rho_p , \quad {\rm with\,\,\,} \mathcal{L}[L]\rho_p = L\rho_p L^\dagger - \frac{1}{2}\{L^\dagger L, \rho_p\},
\end{equation}
where $\Gamma_{dp}$ is the phonon dephasing rate, which is inversely proportional to the coherence time of the phonon mode. In our noisy circuit simulation, we set the coherence time of the phonon mode to be $3 \, ms$. 

To translate this dephasing channel to Kraus operator, we first expand the phonon density matrix $\rho_p(\tau)$ in the Fock-state basis as
\begin{equation}\label{eq:rho-fock-state}
    \rho_p(\tau) = \sum_{m,n} \rho_{nm}(\tau)|n\rangle \langle m |.
\end{equation}
Substituting \eq{rho-fock-state} into \eq{lindblad-phonon-dephasing}, we obtain the analytical solution for the matrix elements:
\begin{align}
    \rho_{nm}(\tau)= e^{-\frac{\Gamma_{dp} \tau}{2}(n-m)^2}\rho_{nm}(0).
\end{align}
This result manifests that the off-diagonal elements of density matrix decay exponentially at a rate proportional to the square of the phonon number difference $(n-m)^2$, while the diagonal elements remain unchanged. This quadratic scaling highlights that superpositions of widely separated Fock states are significantly more susceptible to dephasing noise.

Equivalently, the dynamic evolution of density matrix can be expressed in Kraus form as
\begin{equation}\label{eq:ds-phonon-dephasing}
    \rho_p(\tau)=\int_{-\infty}^{\infty} P(\phi) e^{i\phi \hat{n}} \rho_p(0) e^{-i\phi \hat{n}} d\phi,
\end{equation}
where $\hat{n}=a^\dagger a$ and the weight function $P(\phi) = \frac{1}{\sqrt{2\pi \Gamma_{dp} \tau}} \exp\left(-\frac{\phi^2}{2 \Gamma_{dp} \tau}\right)$ follows a Gaussian distribution. This expression shows that the density matrix at duration time $\tau$ can be interpreted as an ensemble average over random phase fluctuations. 

According to \eq{ds-phonon-dephasing}, the Kraus operators of this dephasing of phonon modes in Fock-state basis can be defined as
\begin{align}
    \hat{E}(\phi) = \sqrt{P(\phi)} e^{i\phi \hat{n}}.
\end{align}
It represents that the dephasing channel is equivalent to an ensemble of unitary rotations in phase space by a random angle $\phi$, with a variance characterized by $\sigma_\phi^2 = \Gamma_{dp} \tau$.
These Kraus operators $\hat{E}(\phi)$ satisfy the completeness relation $\int_{-\infty}^{\infty} \hat{E}^\dagger(\phi) \hat{E}(\phi) \, d\phi = \mathbb{I}$. The dephasing channel of the phonon mode can then be written as $\rho_p(\lambda_{dp})=\mathcal{E}(\rho_{p0})=\int_{-\infty}^{\infty} \hat{E}(\phi) \rho_{p0} \hat{E}^{\dagger}(\phi) d\phi $.

In phase space, this channel effectively "smears" the phase of the phonon mode while preserving its radial distribution (phonon number). This loss of phase information reduces the coherence required for high-fidelity interference, contributing a major noise to the entangling operations like the native Mølmer-Sørensen operation which constitutes the analog gates (\(\mathbf{A}\)) and the CNOT gate in trapped ion devices.

\paragraph{Spin mode}

We next consider the dephasing channel acting on the spin degree of freedom. In contrast to the phonon mode, which evolves in an infinite-dimensional Hilbert space, the spin qubit is encoded in the internal two-level structure of the ion. The dynamics of the spin mode subjected to pure dephasing is governed by the Lindblad equation with the Lindblad operator being the Pauli-$Z$ operator $\sigma^z$:
\begin{equation}\label{eq:lindblad-spin-dephasing}
    \frac{d\rho_s}{d\tau} = \Gamma_{ds} \mathcal{L}[\sigma^z]\rho_s, \quad {\rm with\,\,\,} \mathcal{L}[L]\rho_s = L\rho_s L^\dagger - \frac{1}{2}\{L^\dagger L, \rho_s\},
\end{equation}
where $\Gamma_{ds}$ is dephasing rate of spin mode, inversely proportional to the coherence time. In our noisy circuit simulation, we take the spin coherence time to be $10 \, ms$. 
The solution of \eq{lindblad-spin-dephasing} can be obtained
\begin{equation}
    \rho_s(\tau) = (1-p)\rho_{s0} + p(\sigma^z \rho_{s0} \sigma^z), \quad {\rm with\,\,\,} p = \frac{1 - e^{-2\Gamma_{ds} \tau}}{2},
\end{equation}
where $p$ represents the phase-flip probability and $\rho_{s0}$ is the initial density matrix. This expression shows that the dephasing channel preserves the populations of the spin state while suppressing the off-diagonal coherence elements. Equivalently, the Kraus operators of this dephasing channel of spin mode can be defined as~\cite{nielsen2010quantum} :
\begin{equation}
    \hat{E}_0 = \sqrt{1-p} \,\mathbb{I},\quad \hat{E}_1 = \sqrt{p} \,\sigma^z,
\end{equation}
which satisfy the completeness relation $\hat{E}_0^{\dagger}\hat{E}_0+\hat{E}_1^{\dagger}\hat{E}_1=\mathbb{I}$. The dephasing channel of the spin mode can be written as $\rho_s=\mathcal{E}(\rho_{s0})=\hat{E}_0 \rho_{s0} \hat{E}_0^{\dagger} + \hat{E}_1 \rho_{s0} \hat{E}_1^{\dagger}$.

\subsection{Extrapolation to the zero noise limit}
\label{sec:zne}
To simulate the noisy dynamics of the quantum circuit shown in Fig.1($\mathbf{d}$), we represent the global state by a density matrix $\rho \in \mathcal{H}_{S_1} \otimes \mathcal{H}_{S_2} \otimes \mathcal{H}_{P_s} \otimes \mathcal{H}_{P_a}$, where $\mathcal{H}_{S_i}$ denotes the $i$-th qubit and $\mathcal{H}_{P_{s,a}}$ represent the storage and auxiliary phonon modes, respectively. The Fock space of the phonons is truncated to five, which is sufficient to reduce truncation errors in the subspace encoding the photon occupation states under consideration. We consider two corresponding observables: the storage-mode phonon number, $\hat{n}_s = a_s^\dagger a_s$, which characterizes the survival of the incident photon $\gamma^L\to\gamma^L$, and the two-spin projector onto the doubly excited state, $\hat{P}_{11} = \frac{(I+\sigma_1^z)(I+\sigma_2^z)}{4}$, which characterizes fermion-antifermion pair production $\gamma^L\to f_{\mathbf{p_1}}^1\bar f_{\mathbf{p_1}'}^1$.

In the presence of the noise channels {discussed in the noise-model section of the Supplemental Material}, the physical observables become explicit functions of the effective noise strength $\lambda$, denoted as $O(\lambda) = \text{tr}(\hat{O} \rho(t, \lambda))$, where $\hat{O} \in \{\hat{n}_s, \hat{P}_{11} \}$. Here, $O(\lambda_0)$ represents the expectation value at the baseline noise level $\lambda_0$-corresponding to the natural noise strength in the implemented circuit, while $\lambda > \lambda_0$ denotes an artificially amplified noise level introduced for error-mitigation purposes. Using zero-noise extrapolation (ZNE) as error mitigation method, we can obtain an estimated observed value at the noiseless level~\cite{PhysRevX.8.031027, PhysRevLett.119.180509}. 

To implement ZNE, the circuit noise is scaled from its baseline value \(\lambda_0\) to a set of larger effective values, and the corresponding expectation values \(O(\lambda)\) are evaluated. Such noise scaling can generally be achieved through parameterized noise amplification, or unitary folding in which the total number of gates is increased while preserving the target unitary evolution. In this study, we will adopt the unitary folding for the demonstration of the capability of ZNE.

Since the two-qubit gates (CNOT) and the analog spin-phonon gates ($\mathbf{A}$) are the dominant sources of infidelity, we focus the noise amplification procedure on these operations. Specifically, we replace each unitary gate $U \in \{ \mathbf{A}, \text{CNOT}\}$ by a folded sequence, $U \rightarrow U(U^\dagger U)^n$, where $n \in \{0, 1, 2, \dots\}$. In the ideal noiseless limit, $U^\dagger U = \mathbb{I}$ leaves the evolution of the circuit invariant. In the presence of noise, the folded sequence increases the effective exposure time to noise from $\tau$ to $\tau' = (2n+1)\tau$ and this is equal to amplifying the effective noise strength by a folding factor $\mathcal{N}_s = 2n+1$. Running the circuit at the amplified noise level $\mathcal{N}_s \lambda_0$, we obtain the corresponding noisy expectation value $O (\mathcal{N}_s)$ in terms of the folding factor $\mathcal{N}_s$. 

To understand the dependence of the observable $O (\mathcal{N}_s)$ on the folding factor $\mathcal{N}_s$ under the unitary folding procedure, we can analyze the Kraus representations of the relevant noise channels. 
For the heating rate channel, the variance of the stochastic phase-space displacement $\sigma^2(\tau') = \mathcal{N}_s \Gamma_{heat} \tau$ scales linearly with $\mathcal{N}_s$. As a result, the storage phonon number exhibits a leading linear drift with the folding factor, i.e., $n_s(\mathcal{N}_s) \approx n_s(0) + \mathcal{N}_s \cdot \Gamma_{\rm heat} \tau$. For the phonon and spin dephasing channels, the folding factor enhances the exponential suppression of the off-diagonal elements of the density matrix under the folded sequence. Specifically, phonon dephasing gives $\rho_{nm}^{P_s} \propto \exp[-\mathcal{N}_s \frac{\Gamma_{dp} \tau}{2}(n-m)^2]$, while spin dephasing leads to $\rho_{nm}^{S_i} \propto e^{-2\mathcal{N}_s  \Gamma_{ds} \tau}$. 

Importantly, the combined effect of these noise channels on the circuit dynamics is not simply additive. Instead, it is shaped by the non-commutativity both among the noise channels themselves and between the noise channels and the gate operations. This feature motivates the use of a nonlinear extrapolation model. For example, the heating rate channel ($\mathcal{L}[a^\dagger]$ and $\mathcal{L}[a]$) do not commute with the phonon dephasing channel ($\mathcal{L}[a^\dagger a]$). Consequently, phase-space diffusion and phase randomization become intertwined during the amplified noisy evolution, giving rise to higher-order cross terms in the effective error budget.

\begin{figure}[htbp]
    \centering
    \begin{subfigure}[b]{0.3\textwidth}
        \includegraphics[width=\linewidth]{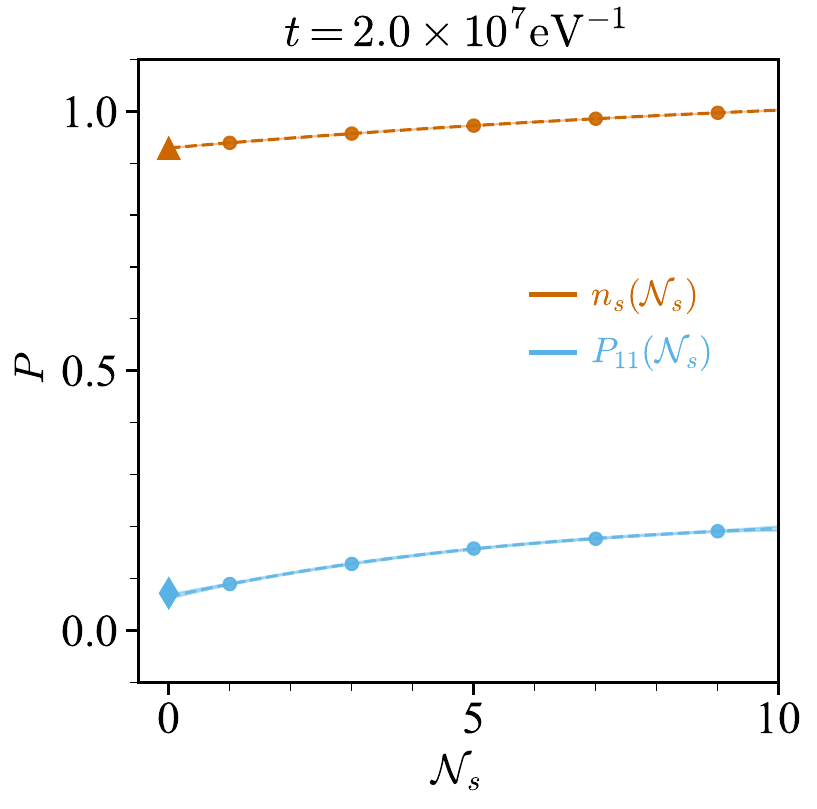}
    \end{subfigure}
    \hfill
    \begin{subfigure}[b]{0.3\textwidth}
        \includegraphics[width=\linewidth]{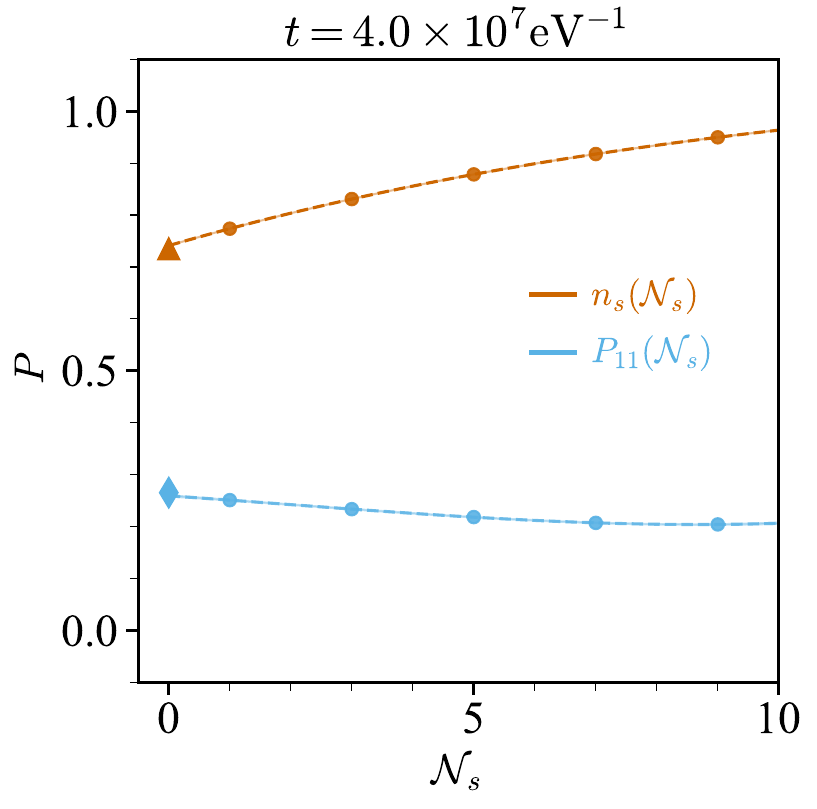}
    \end{subfigure}
    \hfill
    \begin{subfigure}[b]{0.3\textwidth}
        \includegraphics[width=\linewidth]{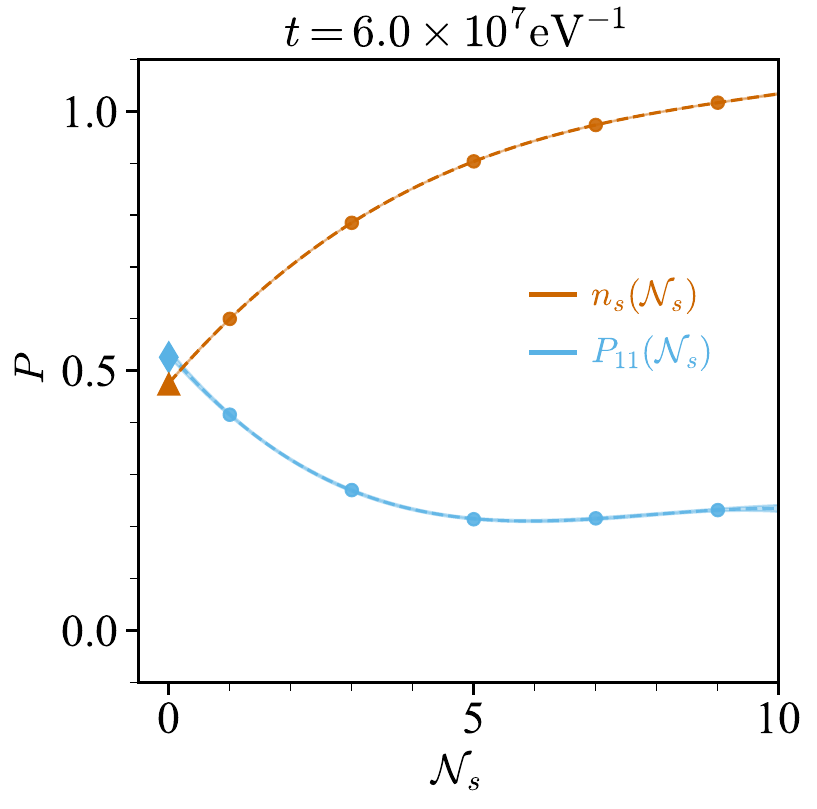}
    \end{subfigure}

    \vspace{0.5em} 

    \begin{subfigure}[b]{0.3\textwidth}
        \includegraphics[width=\linewidth]{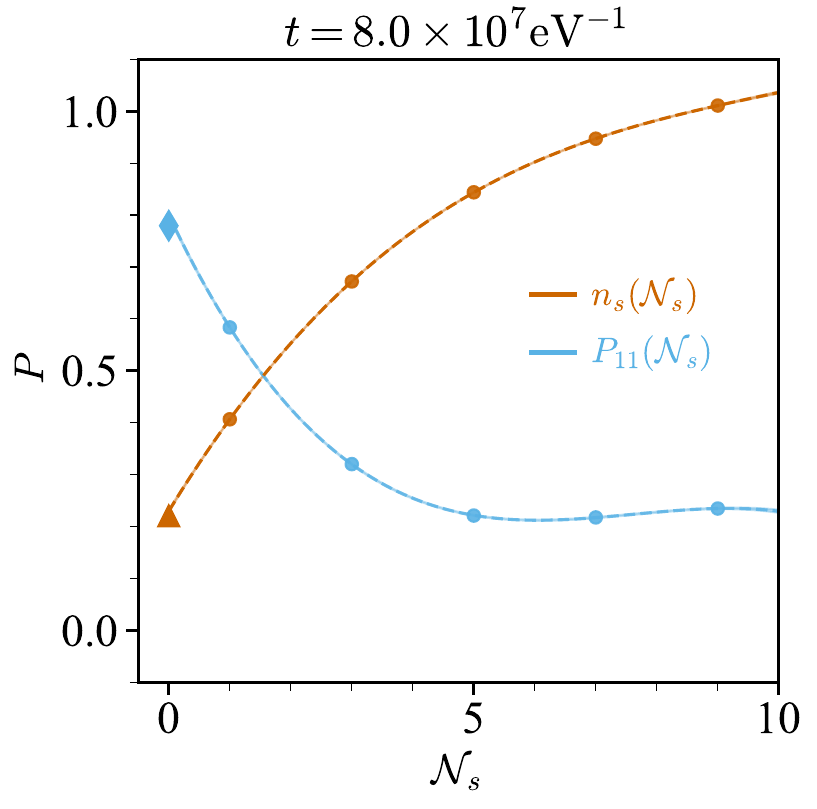}
    \end{subfigure}
    \hfill
    \begin{subfigure}[b]{0.3\textwidth}
        \includegraphics[width=\linewidth]{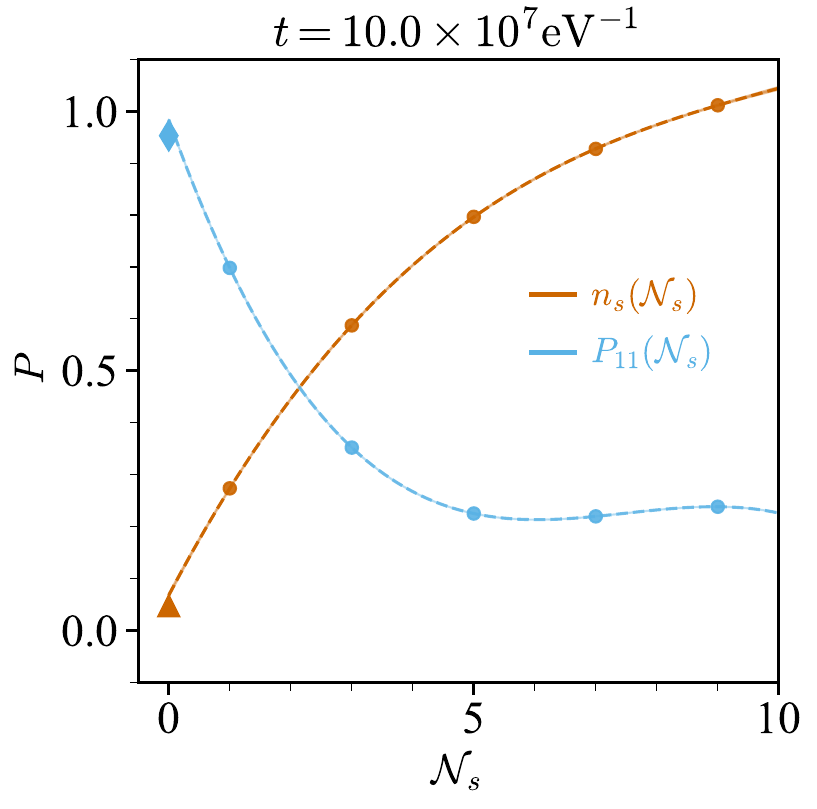}
    \end{subfigure}
    \hfill
    \begin{subfigure}[b]{0.3\textwidth}
        \includegraphics[width=\linewidth]{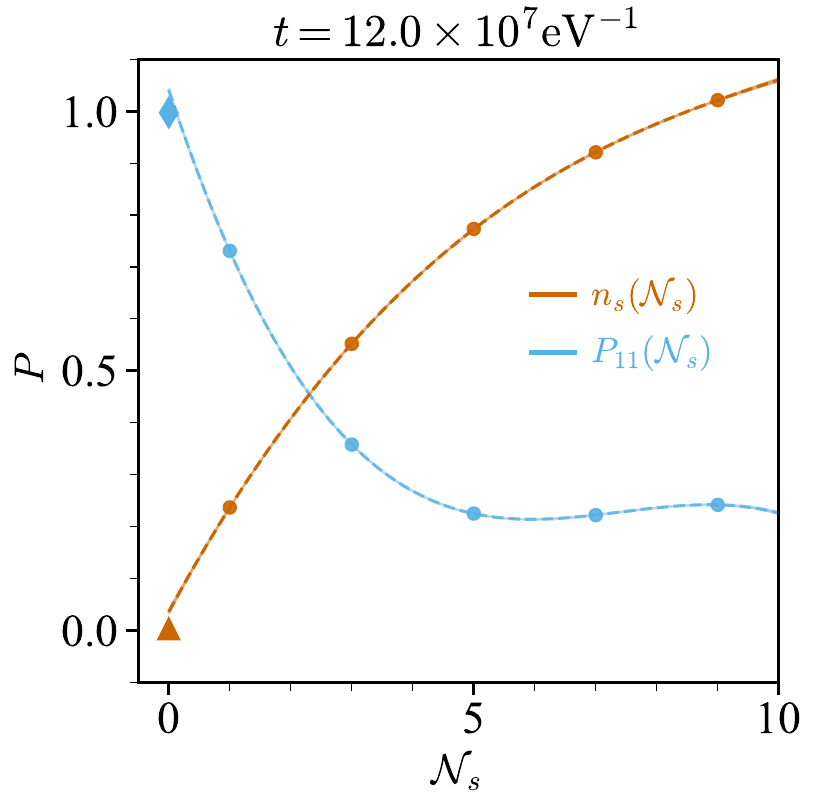}
    \end{subfigure}

    \caption{The survival channel $\gamma^L\to\gamma^L$ (orange, represented by $n_s$) and the pair-production channel $\gamma^L\to f_{\mathbf{p_1}}^1\bar f_{\mathbf{p_1}'}^1$ (blue, represented by $P_{11}$) at different evolution time $t$ from the noisy simulation and the extrapolation to the zero noise limit. Circle markers are simulation date at different folding factor $\mathcal{N}_s$ and dashed lines are the corresponding fitting curves. Blue rhombus (orange triangles) marker show exact $P_{11}$ ($n_s$) values at each time step $t$.}
    \label{fig:zne_deltat}
\end{figure}

Consequently, the observables $O (\mathcal{N}_s)$ are not a simple linear or mono-exponential function. To accurately capture these nonlinear couplings, we utilize third-order polynomial fitting to extract $O_0 =O (\mathcal{N}_s \to 0)$. Thus, the observable $O(\mathcal{N}_s)$ can be expanded in terms of the folding factor to the third-order as:
\begin{equation}
    O(\mathcal{N}_s \lambda_0) = O_0 + \sum_{k=1}^{K=3} a_k (\mathcal{N}_s \lambda_0)^k + \mathcal{O}(\lambda_0^{K+1}).
\end{equation}
To show the effectiveness of our error mitigation strategy, we performed numerical simulations for different evolution times $t$. For each specific time of evolution, we amplified the physical noise by applying gate folding as mentioned. By varying the folding factor $\mathcal{N}_s \in \{1, 3, 5, 7, 9\}$, we obtained a set of noisy observables: the survival storage phonon number $n_s(\mathcal{N}_s)$ and the pair-production probability $P_{11}(\mathcal{N}_s)$. By applying the third-order polynomial function fitting to the $\mathcal{N}_s$-dependent data at each time step $t$, we are able to extract the zero-noise limit $n_s(\mathcal{N}_s\to 0)$ and $P_{11}(\mathcal{N}_s\to 0)$, providing a high-fidelity reconstruction of the ideal quantum dynamics. The results are shown in \fig{zne_deltat}, where circles represent simulation data under different folding factor $\mathcal{N}_s$ and blue (orange) dashed line represents the corresponding fitting curves for $P_{11}$ ($n_s$). Blue rhombus (orange triangles) marker shows exact $P_{11}$ ($n_s$) values at each time step $t$. Extrapolated results at $\mathcal{N}_s = 0$ closely match exact values, confirming the effectiveness of the error mitigation method. 

\end{document}